\documentclass[useAMS,usenatbib]{mn2e}
\usepackage{epsfig}
\usepackage{amssymb}

\newcommand{\msun}{\mbox{M$_{\odot}$}}
\newcommand{\lsun}{\mbox{L$_{\odot}$}}
\newcommand{\kms}{\mbox{km s$^{-1}$}}

\title[The Collision Between The Milky Way And Andromeda]
{The Collision Between The Milky Way And Andromeda}

\author[Cox \& Loeb]
{T. J. Cox\footnotemark[1] and Abraham Loeb\footnotemark[2] \\
	Harvard-Smithsonian Center for Astrophysics,
	60 Garden Street, Cambridge, MA 02138, USA \\
	}

\begin{document}

\maketitle

\begin{abstract}

We use a N--body/hydrodynamic simulation to forecast the future encounter
between the Milky Way and the Andromeda galaxies, given current
observational constraints on their relative distance, relative velocity,
and masses.  Allowing for a comparable amount of diffuse mass to fill the
volume of the Local Group, we find that the two galaxies are likely to
collide in a few billion years - within the Sun's lifetime.  During the the
interaction, there is a chance that the Sun will be pulled away from its
present orbital radius and reside in an extended tidal tail.  The
likelihood for this outcome increases as the merger progresses, and there
is a remote possibility that our Sun will be more tightly bound to
Andromeda than to the Milky Way before the final merger.  Eventually, after
the merger has completed, the Sun is most likely to be scattered to the
outer halo and reside at much larger radii ($>30$~kpc).  The density
profiles of the stars, gas and dark matter in the merger product resemble
those of elliptical galaxies.  Our Local Group model therefore provides a
prototype progenitor of late--forming elliptical galaxies.

\end{abstract}

\begin{keywords}
galaxy:evolution --- galaxies:evolution --- galaxies:formation ---
galaxies:interactions ---  Local Group ---  methods:N-body simulations.
\end{keywords}

\renewcommand{\thefootnote}{\fnsymbol{footnote}}
\footnotetext[1]{tcox@cfa.harvard.edu}
\footnotetext[2]{loeb@cfa.harvard.edu}

\section{Introduction}
\label{sec:intro}

It is well known that the Milky Way (MW) and Andromeda (M31) are the two
largest members of the Local Group of galaxies.  Together with their
$\sim40$ smaller companions, the Milky Way and Andromeda comprise our
galactic neighborhood, and as such, represent the nearest laboratory,
and therefore the most powerful tool, to study the formation and
evolution of galactic structure.

Like most extragalactic groups, the Local Group is very likely to be
decoupled from the cosmological expansion and is now a gravitationally
bound collection of galaxies.  This notion is supported by the observed
relative motion between its two largest galaxies; namely, the Milky Way and
Andromeda are moving toward each other at $\sim120$~\kms \citep{BT}.
Unfortunately, this motion alone does not indicate whether the Local Group
is bound or not.  The unknown magnitude of Andromeda's transverse velocity
adds uncertainty into the present day orbital parameters and therefore the
past and future evolution of the Local Group.

Barring the uncertain transverse velocity of Andromeda, a considerable
amount of information can be inferred about the Local Group provided a
plausible set of assumptions.  Nearly 50 years ago \citet{KW59} pioneered
the ``timing argument,'' in which the Milky Way and Andromeda are assumed
to form within close proximity to each other, during the dense early stages
of the Universe, before they were pulled apart by the general cosmological
expansion.  They have subsequently reversed their path and are approaching
one another owing to their mutual gravitational attraction.  According to
the timing argument, the Milky Way and Andromeda have now traced out nearly
a full period of their orbital motion which is governed by Kepler's laws.
By assuming that the system has no angular momentum, and given the current
separation, velocity of approach, and the age of the Universe, the timing
argument yields estimates for the mass of the Local Group
($>3\times10^{12}$\msun), the semi--major axis of the orbit
($<580$~kpc), and the time of the next close passage ($>4$~Gyr)
\citep[see Sec. 10.2 of ][]{BT}.

While the seminal results of \citet{KW59} were an early indication of the 
large mass--to--light ratio in the Local Group and therefore the presence
of dark matter, they also began a nearly five decade long quest to understand
the past, present, and future of our Local Group.  In particular, a number of
studies have extended the original timing argument by allowing for
various angular momenta, by including more realistic or time--dependent
mass distributions, by adding the effects of mass at scales beyond that
of the Local Group, or testing its validity using numerical simulations
\citep[see, e.g.,][]{Peeb89,FT91,Val93,P94,P01,SF05,Loeb05,LW07,vdM07}.

One of the most intriguing developments stemming from the various studies
of the Local Group is an estimate of the transverse velocity of Andromeda.
By employing the action principle to the motions of galaxies within and
near ($<20$~Mpc) the Local Group, \citet{P01} concluded that the transverse
velocity of Andromeda is less than $200$~\kms.  Using the well measured
transverse velocity of M33 \citep{Bru05} and numerical simulations that tracked
the potential tidal disruption during M33's past encounters with Andromeda,
\citet{Loeb05} found an even smaller estimate, $\sim 100$~\kms, for the
transverse velocity.  While future astrometric observations using 
SIM\footnote{http://planetquest.jpl.nasa.gov/SIM/} and
GAIA\footnote{http://www.sci.esa.int/gaia/}
will be able to accurately measure the proper motion of Andromeda, the
low values favored by these papers suggests that the Local Group is indeed
a gravitationally bound system.

Provided that the Local Group is gravitationally bound, and that the Milky
Way and Andromeda are heading towards each other, one must admit the
possibility that they will eventually interact and merge.  This outcome
appears inevitable given the massive halos of dark matter that likely 
surround the Milky Way and Andromeda.  Numerical experiments have robustly
concluded that dark matter halos can exert significant dynamical friction,
and are sponges that soak up energy and angular momentum leading to a rapid
merger \citep{B88}.

Even though the eventual merger between the Milky Way and Andromeda is
common lore in Astronomy, the merger process has not been addressed by a
comprehensive numerical study.  The one exception is a paper by
\citet{Dub96} that presented a viable model for the Local Group and
numerically simulated the eventual merger between the Milky Way and
Andromeda.  However, \citet{Dub96} utilized this Local Group model and its
numerical evolution to study the production of tidal tails during
such an encounter and the possibility to use the structure of this tidal
material to probe the dark matter potential.  While the study by
\citet{Dub96} provided the first enticing picture of the future encounter
between the Milky Way and Andromeda, \citep[for a more recent and higher
resolution version of this simulation, see][]{Dub06}, it was neither
designed to detail the merger dynamics including intergalactic material,
nor outline the possible outcomes for the dynamics of our Sun, nor quantify
properties of the merger remnant.  In addition, the last decade has
produced a number of improved models for the structure of the Milky Way and
Andromeda as well as the properties of the intragroup medium.

In this paper we quantitatively predict when the interaction and merger of
the Milky Way and Andromeda will likely occur and forecast the probable
dynamics of the Sun during this event.  We achieve this goal by
constructing a model for the Local Group in \S\ref{sec:model} that
satisfies all observational constraints.  We then evolve this model using a
self-consistent N-body/hydrodynamic simulation, as described in
\S\ref{sec:meths}.  The generic properties of the merger, including the
merger timescale, the possible evolution of our Solar System, and
properties of the merger remnant, are outlined in \S\ref{sec:results}.
Finally, we conclude in \S\ref{sec:conc}.

\section{A Model of the Local Group}
\label{sec:model}

The distribution of mass within our Local Group of galaxies has
been a long--standing question in astrophysics.  It is clear that
much of the matter is associated with the two largest galaxies
in the Local Group: the Milky Way and Andromeda.  Moreover, these
two spiral galaxies are likely to be embedded in an ambient medium
of dark matter and gas.

\subsection{The Milky Way and Andromeda}
\label{ssec:mwand}

There are a number of different models for both the Milky Way and Andromeda
galaxies \citep[see, e.g.,][and reference therein] {KZS02,WD05,SBB07}.  These
studies generally enlist a myriad of observational data to infer the
distribution of baryons, while the dark matter, which dominates the
gravitational potential, is set to match distributions extracted from
cosmological N-body simulations \citep[e.g.,][]{NFW96}.  Together, these
models specify the total mass distribution out to the virial radius
($\sim200-300$~kpc).

In our model of the Local Group we start by adopting the models for the
Milky Way and Andromeda favored by \citet{KZS02}.  Within these models, the
baryons are contained entirely within the rotationally supported
exponential disk and central bulge.  These components are then surrounded
by a massive dark--matter halo, which has nearly 20 times the mass as the
baryons, as specified by the mass fractions, $m_b$ and $m_d$, defined as
the bulge and disk mass, respectively, divided by the total mass.  The
exponential disk, of radial disk scale radius $R_d$, also contains a set
fraction $f$ of its mass in collisional gas that can cool and form stars.
Both the bulge and dark halo components are assumed to follow the
\citet{H90} profile.  The bulge scale radius $a$ is fixed to be 20\% of the
radial disk scale radius $R_d$.  The dark--matter profile is defined by its
concentration $c$, spin parameter $\lambda$, and total virial mass
$M_{200}$ and virial circular velocity $V_{200}$ (at the radius $r_{200}$
where the average interior density is 200 times the critical cosmic density
today, $rho_{\rm crit}= 10^{-29}~{\rm g~cm^{-3}}$), which are all listed in
Table~\ref{tab:galps}.  The numerical construction of these models employs
methods commonly used to construct equilibrium disk galaxies \citep[see,
e.g.,][]{H93,SW99,Sp00,Cox06,SdMH05}.

\begin{table}
\begin{center}
\caption{
Properties of the Milky Way (MW) and Andromeda (M31) models
used in the work (see text for definitions).
}
\begin{tabular}{lcc}
\hline
Property & MW & Andromeda \\
\hline
$V_{200}$ (\kms) &   145  & 170  \\
$M_{200}$ ($10^{12}$\msun) &   1.0  & 1.6  \\
$c$              &   12 & 12 \\
$\lambda$        &  0.031 & 0.036  \\
$m_d$            &  0.041 & 0.044  \\
$R_d$            &  2.2 & 3.6  \\
$f$ &  0.3   & 0.3  \\
$m_b$            &  0.008 & 0.012  \\
$a$            &  0.4 & 0.7  \\
N$_{dm}$       & 475,500 &  755,200 \\
N$_{disk}$       & 14,350 &  24,640 \\
N$_{gas}$       & 6,150 &  10,560 \\
N$_{bulge}$       & 4,000 &  9,600 \\
\hline
\end{tabular}
\label{tab:galps}
\end{center}
\end{table}

\subsection{The Orbit}
\label{ssec:mwandorb}

Given the adopted parameters of the two largest galaxies in the Local
Group, we must now define their orbital parameters and any ambient medium
in which the system will be embedded.  There are a few empirical
constraints that must be considered.  First, at the current epoch, the
separation between the Milky Way and Andromeda is 780~kpc
\citep{McC05,Rib05}.  Second, the Milky Way and Andromeda are approaching
each other at a radial speed of 120~\kms, assuming a local circular speed
of 220~\kms \citep[see Sec. 10.2 of][]{BT}.  These observational facts
tightly constrain any dynamical model of the Local Group since their
fractional error bars are estimated to be less than $\sim5$\%.

Less well constrained is the current estimate for the transverse velocity
of Andromeda.  As mentioned in \S\ref{sec:intro}, speeds of $<200$~\kms are
favored by recent models \citep{P01,Loeb05,vdM07}, but depend upon
assumptions regarding the distribution of mass within the Local Group and
its initial state.  We will therefore gauge the success of our Local Group
model by its ability to reproduce the above three observations, but it
should be kept in mind that the first two observations have significantly
less leeway than the third.

Constraints on the spin orientation of both the Milky Way and Andromeda
with respect to the orbital plane of the merger originate from the current
position and orientation of Andromeda in the night sky.  While these
details are necessary for a complete model of the Local Group, they do not
significantly influence the timing of the merger between the Milky Way and
Andromeda.  The spin orientation will, however, affect the disk morphology
during the merger, a detail that will be addressed when we attempt to track
the possible dynamical fate of our Sun in \S\ref{ssec:earth}.

\subsection{The Local Intragroup Medium}
\label{ssec:lgmed}

One plausible starting model for the Local Group is to follow the logic
originally employed by the timing argument \citep{KW59}, i.e., that the
mass of the Local Group is entirely contained within the Milky Way and
Andromeda and their motion is a simple two body problem governed by
Kepler's equations.  In practice, however, most of the recent
implementations of the timing argument \citep[see, e.g.,][] {BT,FT91,LW07}
generally yield masses for the Local Group ($>3\times10^{12}$~\msun) that
exceed the total masses in our Milky Way and Andromeda models
($2.6\times10^{12}$~\msun).  This discrepancy suggests that the Milky Way
and Andromeda do not contain the entire quantity of mass in the Local Group
and are instead the most massive concentrations of mass within a larger
all--encompassing medium.  This point of view is very natural in a
cosmological context where galaxies are not isolated islands, but rather
mountain peaks within a vast continent of land.

\begin{figure*}
\begin{center}
\resizebox{14.0cm}{!}{\includegraphics{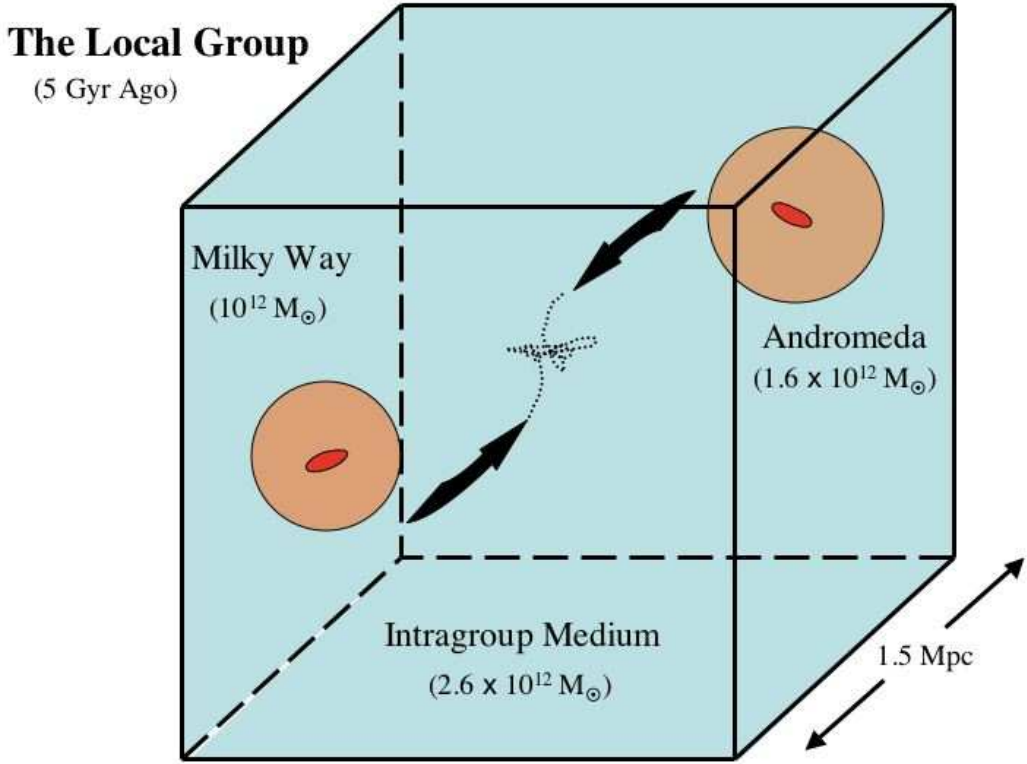}}\\
\caption{
Sketch of the initial configuration of our Local Group model, which
consists of the Milky Way and Andromeda embedded in a diffuse,
constant--density intragroup medium of equivalent mass.
See the text in \S\ref{sec:model} and Table 1 for more details.
\label{fig:lgmodel}}
\end{center}
\end{figure*}

For these reasons our Local Group model supplements the Milky Way and
Andromeda galaxy models with a diffuse and extended intragroup medium, as
schematically depicted in Figure~\ref{fig:lgmodel}.  Put within a
cosmological framework this initial configuration may represent a point in
time when the entire Local Group has decoupled from the general Hubble flow
and is in the process of collapsing to become a virialized structure.

For simplicity, we assume that the Local Group medium is initially a
constant density cube of 1.5 Mpc on a side composed of both dark matter and
gas.  The total diffuse mass within this cube is set equal to the total
mass of the two galaxy halos today, $2.6\times 10^{12}$~\msun, yielding a
net mass interior to the MW/M31 orbit that is consistent with the timing
argument.  Our choice is motivated by other cosmological simulations
\citep{Gao04a} which indicate that a substantial fraction of the dark
matter is likely to be diffuse and resides in between virialized halos
(within the unvirialized Local Group) at the initial time of our simulation
$\sim5$~Gyr ago.
\footnote{The initial scale of the Local group in our simulation is an
order of magnitude larger than the virial radius $r_{200}$ of the Milky Way
or Andromeda galaxies. If one were to extend the \citet{NFW97} density
profile of the envelope around each galaxy (with an asymptotic radial
dependence of $r^{-3}$) out to our initial Local Group scale, then one
would roughly double the mass found within the virial radius of each
halo. The added mass in that case would be similar to the amount we indeed
assume for the intragroup medium.}

We also postulate that the Local Group
volume contains close to the cosmic mean value of baryons, namely 16\%
\citep{Sp03}.  Since the Milky Way and Andromeda galaxy models only include
baryons in the galactic disk and bulge components, they are far short of
this value.  We therefore set the Local Group medium to be 20\% primordial
gas, by mass, so that the entire region approaches the cosmic mean value.
The gaseous component, like the dark matter, is initialized to have a
constant density, and its temperature is fixed to be $3\times10^5$~K,
consistent with estimates for the warm--hot intergalactic medium at
comparable overdensities from cosmological simulations \citep[see, e.g.,
Fig.~6 in][]{Dav01} and with observations of apparent RAM pressure
stripping in the Local Group dwarf galaxy Pegasus \citep{McC07}.

We note that the initial gas temperature is far below the virialized
temperature of the Local Group, however this assumption is consistent with
the post--turn around, and pre--virialized initial conditions that we
adopt.  The constant density is also not in line with the assumed growth of
the initial stellar disks from a virialized gaseous halo \citet[see,
e.g.,][ and references therein]{MMW98}.  Within a halo dynamical time, some
of the initial intragroup gas is expected be accreted by each galaxy, and
most of the remaining intragroup gas will end up being shock--heated during
the collapse and relaxation of the Local Group. The dynamics of the merger
between the Milky Way and Andromeda is dominated by the distribution of
dark matter and stellar disks, and so the basic results of our simulation,
involving for example the timing of the merger, are not directly affected
by the initial gas temperature and density distribution.

Since the total mass of the diffuse medium is equivalent to the two
galaxies, the two components are represented with an equivalent number of
particles; 1.3 million (1.04 million dark matter, and 260,000 gas).  With
these assumptions, the overdensity of our initial Local Group model is
$\delta_{\rm LG} = \rho_{\rm LG}/\rho_{\rm crit} \approx 10$.

\subsection{Other Considerations}
\label{ssec:other}

Unfortunately, once the Local Group contains this extended mass
distribution the relative motion of the Milky Way and Andromeda no
longer becomes a trivial application of Kepler's laws.  The diffuse
mass will steadily extract orbital energy and angular momentum owing
to dynamical friction, and the deep potential wells of the galaxies
will slowly accrete diffuse matter.  Both of these effects act to 
gradually extract energy from the binary system, hardening its
orbit, and accelerating the merging process.

Owing to the complicated nature of the orbit, and its deviation from
simple two--body motion, we are left with a fair degree of ambiguity
regarding the initial state of our Local Group model.  Because the 
configuration envisaged by Figure~\ref{fig:lgmodel} resembles the cosmological
collapse of an overdense region of the Universe, a natural starting
point is when this fluctuation decoupled from the general cosmological
expansion, turned around, and began to collapse under its own self gravity.
Like the original timing argument, we then presume that this scenario also
occurred for the Milky Way and Andromeda as well, i.e., we initialize their
radial velocity to be zero as if they have just turned around and are now
about to begin their gravitational collapse.

While some insight about the value of the turn around radius can be gained
from an application of the timing argument, in practice we find that the
position and velocity of the Milky Way and Andromeda quickly deviated from
simple two--body motion owing to the diffuse intragroup medium.  In
general, the velocities quickly grew larger than the original orbit
specified and the direction became highly radial.  We therefore adopt a
trial and error approach where we build an initial model, run it until the
Milky Way and Andromeda are separated by 780~kpc (as this is their current
separation), and inspect the relative velocity between the two galaxies to
assess the validity of the model.

Employing this procedure results in a number of models that are all honest
representations of the Local Group at the present time.  In practice, we
find a trade off between the initial separation and the initial orbital
energy such that models which start with a larger separation require a
smaller initial eccentricity.  In all cases, the apparent orbit of the two
galaxies becomes (or was to begin with) nearly parabolic at the present
time, consistent with recent estimates of the Milky Way--Andromeda orbit
\citep[see, e.g.,][] {vdM07}.  While the rest of this paper will primarily
present the results of one particular model, we will also show that all
models yield similar estimates for the eventual merger between the Milky
Way and Andromeda.  We will argue in \S\ref{ssec:time} that this
convergence results naturally from our assumed intragroup medium.

The model we choose to focus upon begins with an initial separation of 1.3
Mpc, and initializes the Milky Way and Andromeda on an eccentric orbit 
$\epsilon=0.494$, with a distance at perigalacticon of 450~kpc.  With this
orbit the initial angular velocity is 65~\kms, which could likely originate
from tidal torques \citep{GT78,RLB89}.  This particular model begins with
the largest separation of all our models and therefore may be the best
representation of the evolution of the Local Group since its decoupling from
the universal expansion.  Since this model tracks the Local Group the farthest
into the past, the intragroup medium also has a significant amount of time to
react to the two galaxies and therefore is likely to be the most insensitive
to its initial configuration.

\section{Numerical Methods}
\label{sec:meths}

To simulate the evolution of our Local Group and in particular the
interaction between the Milky Way and Andromeda use the publically
available N-body/hydrodynamic code {\small GADGET2} \citep{SpGad2}.  This
version of the code employs the ``conservative--entropy'' formulation of
Smoothed Particle Hydrodynamics \citep[SPH,][]{SHEnt} that conserves both
energy and entropy \citep[unlike earlier versions of SPH; see
e.g.,][]{H93sph}, while improving shock-capturing.  We assume that the gas
is of primordial composition, and include the effects of radiative cooling.

Star formation and its associated feedback are both included in a
manner very similar to that described in \citet{Cox06}.  As is
commonly assumed, stars are stochastically formed at a rate
determined by the SPH gas density \citep[see, e.g.][]{Kz92,SH03,SdMH05,
Cox06} with an efficiency set to match the observed correlation 
between star formation and gas density \citep{Kenn98}.

Feedback from stellar winds and supernovae is treated in a very simplistic
manner, namely the SPH particles that have sufficient density to form
stars are fixed to have an effective temperature of $10^5$ K.  This
methodology is similar in principle to most of the currently favored models
for feedback \citep[see, e.g.,][]{Sp00,SH03, Stin06,Cox06}, and is easy to
implement.  Since the focus of this work is the large--scale evolution of
the Local Group and the generic dynamics of the collision between the Milky
Way and Andromeda, the detailed treatment of the inter--stellar medium does
not influence our primary conclusions.

Numerical resolution is a significant consideration for any computational
problem.  For our purposes here, we require sufficient resolution to
reliably follow the interaction and merger of the Milky Way and Andromeda,
while maintaining the ability to perform a number of simulations with the
available computational resources.  These considerations motivated the
particle number choices outlined in \S\ref{sec:model}.  Given the large
number of components in these simulations (stellar disks, dark halos, and
intragroup medium) and the desire to reduce two--body effects, we required
all particles to have an identical mass of $2\times 10^{7}$\msun and we
employed a universal gravitational softening length of 150~pc.  To test the
sensitivity of the results to these parameter choices, we also ran a higher
resolution version of one model with 30 times the baryonic disk mass
resolution (and therefore number of particles) and 2 times the resolution
of the dark matter and intragroup medium.  In this case we also decreased
the gravitational softening length of the baryons by a factor of 3, and
increased that of the dark matter by a factor of 3.  While this test
yielded much better resolution of the stellar disks and in particular the
tidal material, the general merger dynamics were identical to the low
resolution version.

\section{The Collision Between The Milky Way and Andromeda}
\label{sec:results}

\subsection{Generic Features of the Merger}
\label{ssec:generic}

\begin{figure*}
\begin{center}
\resizebox{16.0cm}{!}{\includegraphics{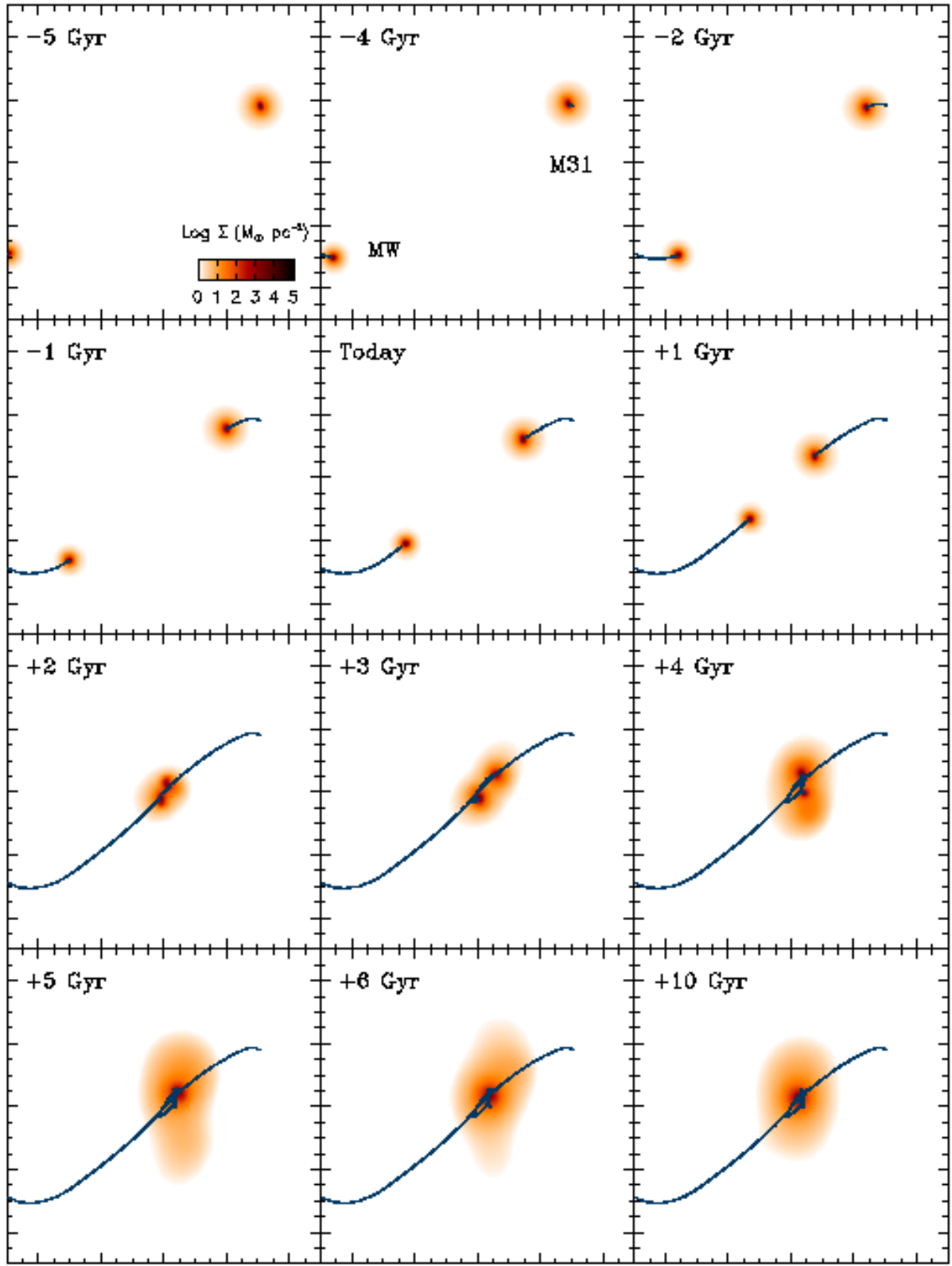}}\\
\caption{ Time sequence of the projected stellar density, shown in
red-scale, for the merger of the Milky Way (MW) and Andromeda (M31).
Andromeda is the larger of the two galaxies and begins the simulation in
the upper-right.  The Milky Way begins on the edge of the image in the
lower-left.  Each panel is 1.5~Mpc square, and the simulation time, in Gyr
relative to today, appear on the top-left label of each panel.  The
trajectories of the Milky Way and Andromeda are depicted by the blue solid
lines.  }
\label{fig:starimages}
\end{center}
\end{figure*}

\begin{figure*}
\begin{center}
\resizebox{16.0cm}{!}{\includegraphics{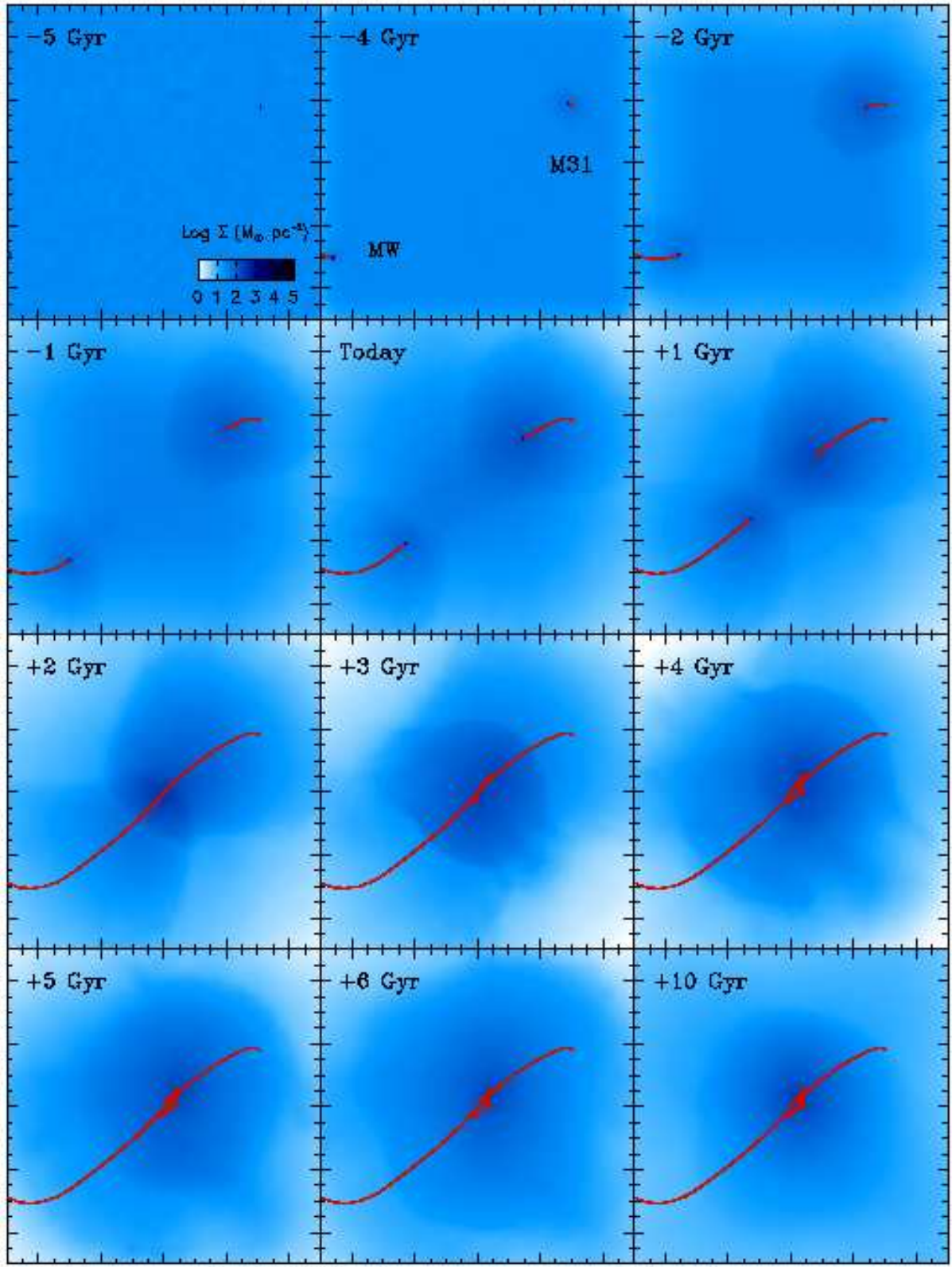}}\\
\caption{ Same as Figure~\ref{fig:starimages} only the projected gas
density is shown.  }
\label{fig:gasimages}
\end{center}
\end{figure*}

\begin{figure*}
\begin{center}
\resizebox{16.0cm}{!}{\includegraphics{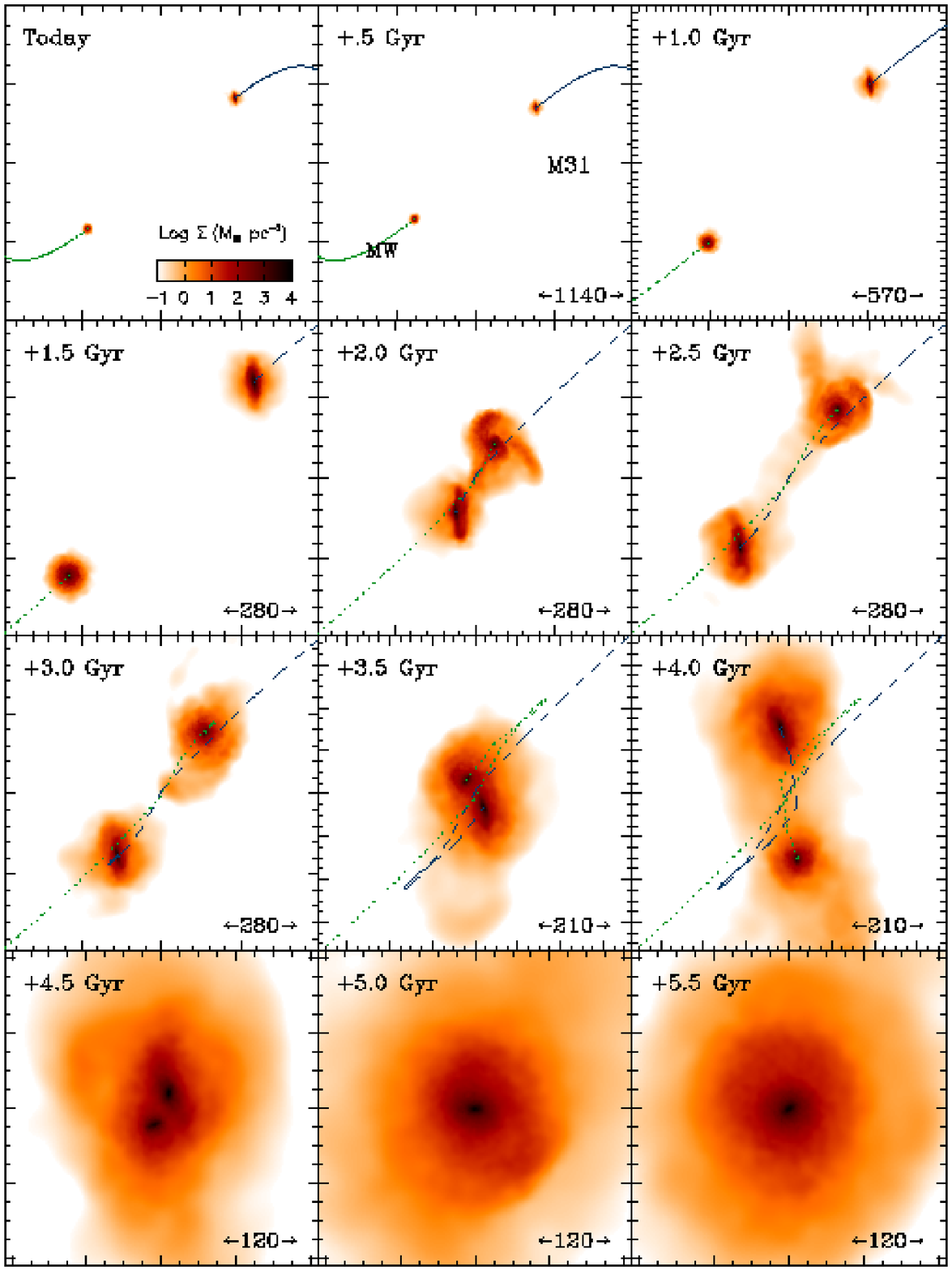}}\\
\caption{ Time sequence of the projected surface mass density of stars,
during the final merger between Andromeda and the Milky Way.  Panels have
varying scales as specified by the label, in kpc, on the lower--right of
each panel. The top--left panel is identical to the ``Today'' panel in
Figure~\ref{fig:starimages}.  The simulation time, with respect to today,
is shown in the top--left of each panel.}
\label{fig:zoomimage}
\end{center}
\end{figure*}

\begin{figure}
\begin{center}
\resizebox{8.0cm}{!}{\includegraphics{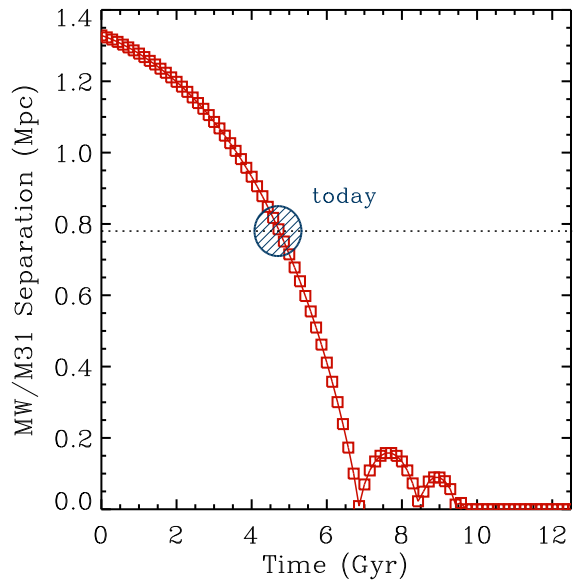}}\\
\caption{ The separation between the centers of Andromeda and the Milky Way
during the course of their merger.  The current separation of $\sim780$~kpc
is shown with a horizontal dashed line and occurs at $T\approx4.7$~Gyr.  }
\label{fig:csep}
\end{center}
\end{figure}

\begin{figure}
\begin{center}
\resizebox{8.0cm}{!}{\includegraphics{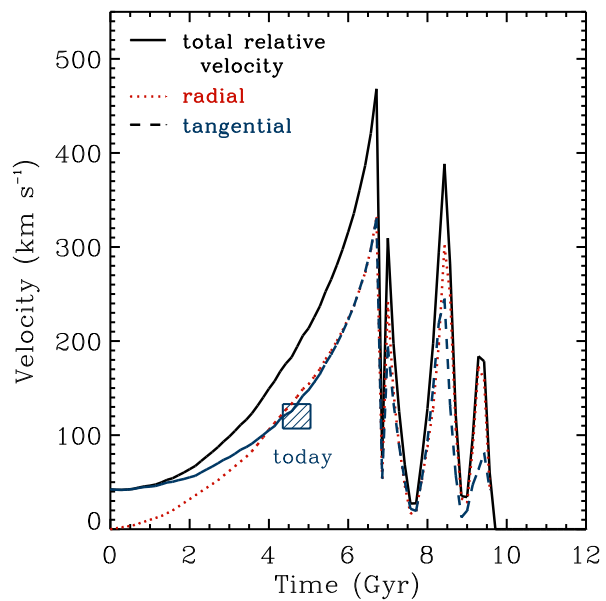}}\\
\caption{ The relative velocity between the centers of Andromeda and the
Milky Way galaxies during the course of their merger.  The present
state of the Local Group occurs at $T\approx4.7$~Gyr. }
\label{fig:crelvel}
\end{center}
\end{figure}

In Figures~\ref{fig:starimages}~through~\ref{fig:crelvel} we present the
basic properties of the dynamical evolution of our Local Group, from 5~Gyr
in the past and until 10~Gyr into the future, beyond the merger time
between the Milky Way and Andromeda.  Most of the features present in these
figures are generic to binary galaxy interactions, and have been described
in great detail by prior studies \citep[see, e.g.,][]{TT72,BH91,
BH92rev,MH96,Cox06}. However, we will review some of the details that are
particularly relevant to the Local Group, and subsequently highlight the
unique status of our own Sun which will be a participant in this galaxy
interaction. The future evolution of structures beyond the local group was
simulated elsewhere \citep{NL03,NL04,Bus03,Bus05}.

To begin, Figures~\ref{fig:starimages}~and~\ref{fig:gasimages} present
the entire evolution of the Local Group from the point of view of a
distant observer.  These images begin at the start of our simulation,
when the Milky Way and Andromeda are separated by 1.3~Mpc, and include
the present state of the Local Group (labeled ``Today'')
and the eventual merger of the Milky Way and Andromeda.  As a guide to
the eye, each panel includes the trajectory of both the Milky Way and
Andromeda.

Shown in Figure~\ref{fig:starimages} is the evolution of the stellar
component, which in our simulation only has contributions from the Milky
Way and Andromeda as we ignore any structure smaller than the two largest
galaxies in the Local Group.

Figure~\ref{fig:gasimages} presents the projected gas distribution during
the interaction, with panels shown at the same times as in
Figure~\ref{fig:starimages}.  Here the color--scale has been stretched to
emphasize the abundant quantity of low--density gas that is spread
throughout the local group.  The initial condition of our Local Group model
assumes a uniform distribution of warm gas, however the gas quickly
responds to the non--uniform potential.  In particular gas is accreted and
shocked to form a hydrostatic halo of warm gas around the Milky Way and
Andromeda galaxies.  The gas distribution is also clearly affected by the
interaction itself, as shocks develop once the galaxy halos begin to
interpenetrate at close separation.

While a detailed analysis of the diffuse gaseous intragroup medium is
beyond the primary focus of our work, the basic properties of this
component appear to be consistent with observations.  In particular, our
model predicts that, at the present state, the intragroup medium is
predominantly warm $10^5-10^6$~K, has a fairly low density, $10^{-4} -
10^{-6}$~cm$^{-3}$, and fills the entire volume that we simulate.  We note
this medium is also expected to extend to larger scales into what has been
termed the ``Warm-Hot Intergalactic Medium" \citep{CO99,HGM98,Dav01}.
While the presence of our intragroup medium is consistent with all current
data \citep{Oso02}. Owing to the difficulty in directly observing gas in
this state, most evidence for its existence comes from indirect means.  For
example, $Chandra$ and $FUSE$ observations of $z\sim0$ Oxygen and Neon
absorption along numerous sight lines suggests the presence of a local,
volume--filling, diffuse, warm medium \citep{Nic02,Nic03,Sem03,Sav03},
although detailed analysis of the ionization states suggest that this is a
complex multi--phase medium that our simulation does not have the
resolution to model.

While Figure~\ref{fig:starimages} presented the dynamical evolution of
the stellar mass on large scales, this vantage point makes it difficult
to distinguish the tell-tale signs of a galaxy merger.  We therefore
zoom into the central regions of the Local Group and specifically show the
merger between the Milky Way and Andromeda in Figure~\ref{fig:zoomimage}.
From this viewpoint the classic signatures of a galaxy interaction, such 
as tidal tails, plumes, and shells are clearly evident.

The physical separation between the Milky Way and Andromeda is presented in
Figure~\ref{fig:csep}.  The current state of the Local Group occurs
$\sim5$~Gyr after the start of our simulation. For the standard set of
cosmological parameters \citep{Teg06short}, this implies that we initiate the
simulation at a redshift of $z\approx 0.5$, around the time when the Sun
was born in the Milky Way disk\footnote{Note that extending the simulation
to significantly earlier times is not adequate since stellar ages imply
that the two galactic disks (and presumably their halos) have not been
fully assembled at $z\ga2$ \citep{Wyse07}.}.  Figure~\ref{fig:csep} also
clearly shows the prediction that our models makes for the future collision
between the Milky Way and Andromeda.  Their first close passage will occur
less than 2~Gyr from the present, and the centers are fully coalesced in
less than 5~Gyr.  These time scales are comparable to the lifetime of our
Sun \citep{SBK93} and admit the possibility that an observer
in the Solar System will witness some (or all) of the galaxy collision.
We will return to this possibility in \S\ref{ssec:earth}.

Finally, we present the relative velocity between the Milky Way and
Andromeda in Figure~\ref{fig:crelvel}.  As in Figure~\ref{fig:csep}
we clearly delineate the present state of the Local Group with a
hatched region whose size corresponds to the errors estimated with the
velocity measurements.  We note that the velocities presented are
relative to each galaxies center of mass, and do not correct for any
motion relative to that.

Now that the general features of our Local Group model have been
outlined, we next explore the validity of our model, the merger
dynamics, the star formation during the interaction, and the
properties of the merger remnant.

\subsection{The Current State of the Local Group}
\label{ssec:present}

Although our model for the Local Group presents its past, current, and
future evolution, the only way to test its validity is by comparing it to
the empirical data on its present-day state.  As we focus primarily on the
evolution of the two largest galaxies in the Local Group, the Milky Way and
Andromeda, it is only the relative separation and motion of these two
galaxies that can be compared to data.  As mentioned in both
\S\ref{sec:intro} and \S\ref{sec:results}, the separation and
line--of--sight velocity of the Milky Way and Andromeda are currently
measured to be 780~kpc \citep[][and references therein]{McC05, Rib05}, and
-120~\kms \citep{BT}, respectively.  The proper motion of Andromeda
perpendicular to our line of sight is less well constrained, but current
estimates suggest that it is $<200$~\kms \citep{P01,Loeb05}.

Figures~\ref{fig:csep} and \ref{fig:crelvel} present the direct comparisons
between the observational constraints and our model, and clearly
demonstrate that it is viable.  In particular, at a time of $4.7$~Gyr after
the start of the simulation (close to the present cosmic time) the
separation between the Milky Way and Andromeda is 780~kpc, while the
relative radial velocity is 135~\kms and the tangential velocity is
132~\kms.  While the tangential velocity is well within the limits
currently favored, the line--of--sight velocity is slightly larger than
(but within 2.0~$\sigma$ of) the observations.  Even though a number of our
model assumptions can be manipulated to reduce these values, e.g., the
initial separation and eccentricity of the fiducial orbit, we note that our
model is a good fit to the velocity when the separation between the Milky
Way and Andromeda is larger than 780~kpc.  Given the observational
uncertainties in these values we feel that there are likely to be a large
number of models that can simultaneously fit all the data within its
2$\sigma$ error bars.

\subsubsection{An Ensemble of Models}
\label{ssec:ensemble}

\begin{figure}
\begin{center}
\resizebox{8.0cm}{!}{\includegraphics{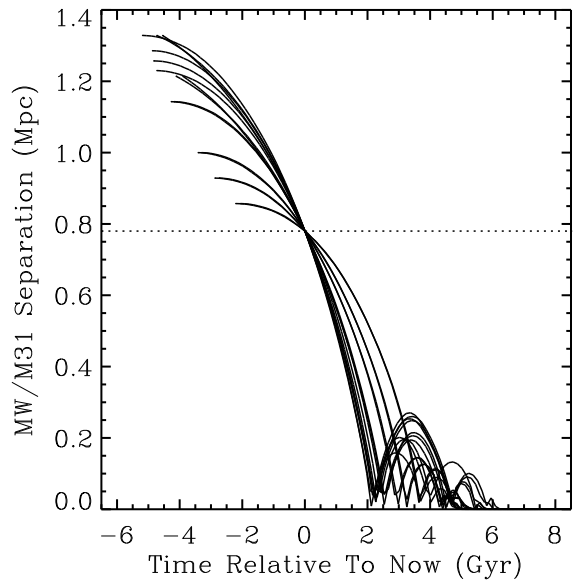}}\\
\caption{ The separation between the centers of Andromeda and the Milky Way
during the course of their merger for a large ensemble of Local Group
models.  The current separation of $\sim780$~kpc is shown with a horizontal
dashed line and all models are normalized to this particular time. }
\label{fig:csep_ensemble}
\end{center}
\end{figure}

As mentioned in \S\ref{ssec:other}, the model described up to this point is
one of twenty Local Group models that we have simulated.  Since the primary
conclusion of this paper, involving the timescale for the eventual merger
between the Milky Way and Andromeda, may be influenced by any number of our
model assumptions, we explicitly show the separation between the Milky Way
and Andromeda and therefore the time of the merger, for all of our models
in Figure~\ref{fig:csep_ensemble}.

Figure~\ref{fig:csep_ensemble} demonstrates several interesting features.
Nearly all of the models provide a similar outcome for time when the two
galaxies make their first passage (ensemble average and standard deviation
are $T=2.8\pm0.5$~Gyr). The same holds true for the final merger time
($T=5.4\pm0.4$~Gyr).  We also note that these average values are slightly
larger than one provided by the model described up to this point. This
highlights a general trend, namely the mergers that start with a larger
separation and have to traverse a longer path through the intragroup
medium, usually (but not always) have a quicker merger dynamics.
Regardless of the relatively small differences between merger times in
these models, they are all completely coalesced by 6.2~Gyr from today.  In
the following section we will argue that this result is a direct byproduct
of our inclusion of an intragroup medium.

\subsection{Merger Dynamics}
\label{ssec:mergerdyn}

\subsubsection{Timescale}
\label{ssec:time}

One of the most intriguing characteristics of our Local Group model is the
relatively quick timescale for the interaction and merger between the
Milky Way and Andromeda.  As stated in the previous section, their first
close passage will occur in less than 2~Gyr, and the final coalescence will
occur in less than 5~Gyr.  While the following section will specifically
address the possible dynamics of the Sun during the merger, we show here
that the cause for the short merger timescale is the intragroup medium.

\begin{figure}
\begin{center}
\resizebox{8.0cm}{!}{\includegraphics{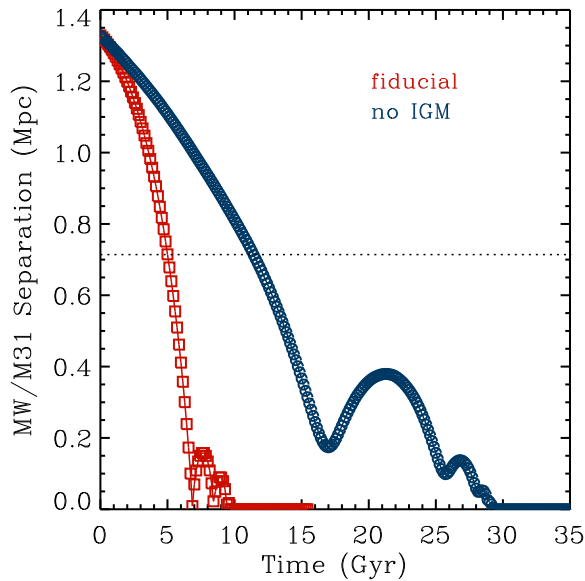}}\\
\caption{ The separation between Andromeda and the Milky Way during the
course of their merger. Our fiducial model in red is compared to a case
with no intragroup medium (dark matter$+$gas) in blue. Dynamical friction
on the diffuse medium shortens the MW/M31 coalescence time by a factor of a
few. The present time is 5 Gyr after the start of the simulation. }
\label{fig:csepboth}
\end{center}
\end{figure}

Our model for the Local Group assumes that the Milky Way and Andromeda are
embedded in a diffuse medium of dark matter and gas.  The medium itself
(minus the evolved galaxies) is unvirialized and of low overdensity
($\delta\sim5$) corresponding to a region of the Universe that has
decoupled from the Hubble flow and started its evolution toward a collapsed
virialized state.  This medium exerts dynamical friction on the Milky Way
and Andromeda and speeds up the merger dynamics by soaking orbital energy
and angular momentum.  This is shown explicitly in
Figure~\ref{fig:csepboth}, which compares the Milky Way -- Andromeda
separation in our fiducial model to an identical model without a diffuse
medium.  Without the intragroup medium, the merger timescale is nearly
three times longer than with it.

The effect of the intragroup medium should scale similarly to the standard
(Chandrasekhar) formula for dynamical friction \citep[Eq. 7-18]{BT}, in
which the deceleration of a massive object is proportional to the
background matter density, $d{\rm v}/dt \propto \rho$.  Therefore, the rate
at which angular momentum is extracted from the orbit depends on the
assumed intragroup medium density, a quantity which is poorly constrained
observationally.  Once the dark matter halos begin to interpenetrate, i.e.,
when the Milky Way -- Andromeda separation is $\sim100$~kpc, the merger
completes relatively quickly because the dark matter halos dominate over
the background density.

This last point is particularly relevant to simulations of binary galaxy
mergers, which typically omit any background overdensity.  While this
omission does not significantly alter the dynamical friction estimates once
the halos overlap and therefore dominate the background density, it may
still change the distribution of orbits in high density environments where
the galaxies traverse through significant overdensities prior to their
interaction event.  In this sense, the merger timescales extracted from
binary merger simulations may be considered an upper limit that is most
applicable to galaxies in the field.

\subsubsection{The Fate of our Solar System}
\label{ssec:earth}

\begin{figure*}
\begin{center}
\resizebox{4.5cm}{!}{\includegraphics{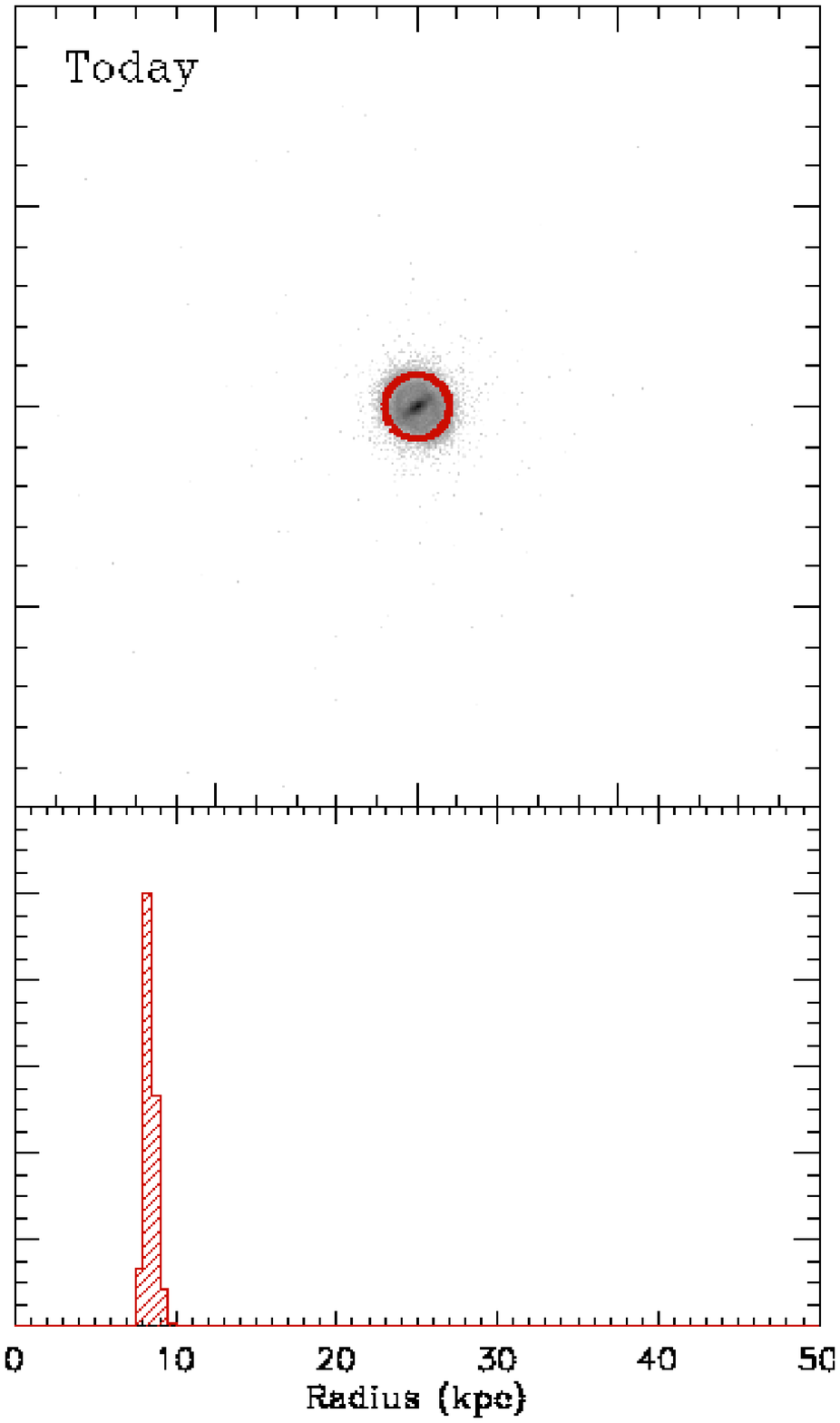}}
\resizebox{4.5cm}{!}{\includegraphics{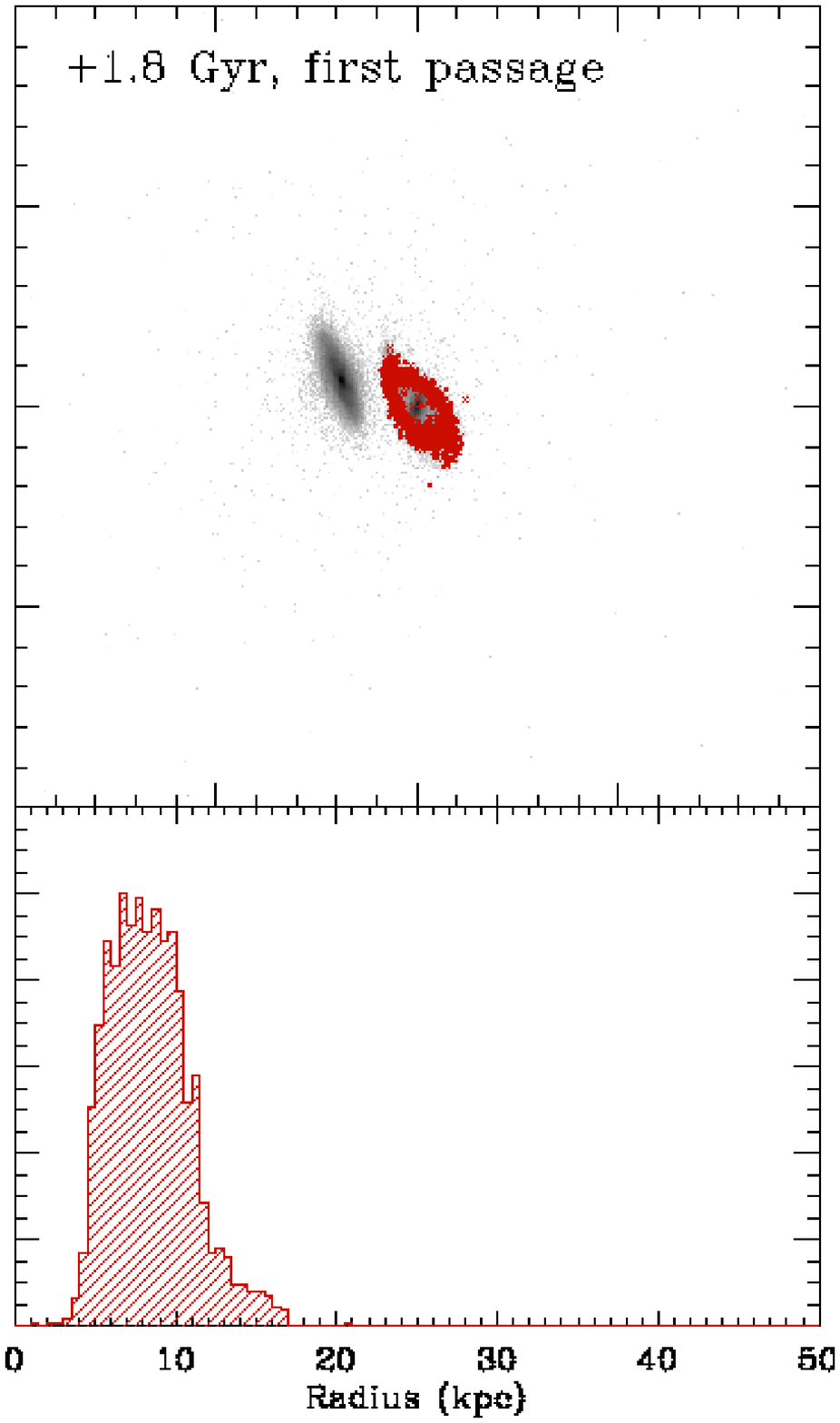}}
\resizebox{4.5cm}{!}{\includegraphics{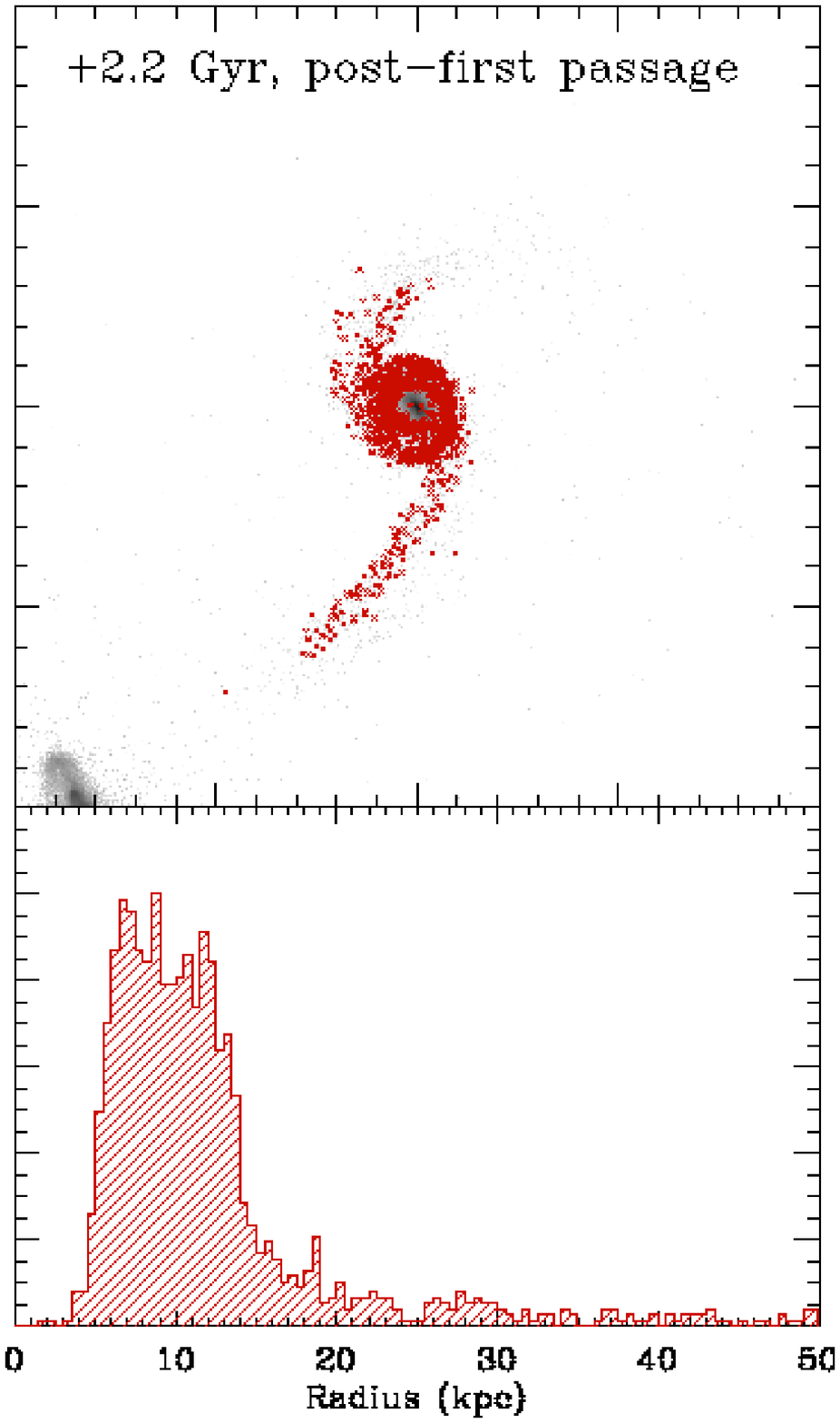}}\\
\resizebox{4.5cm}{!}{\includegraphics{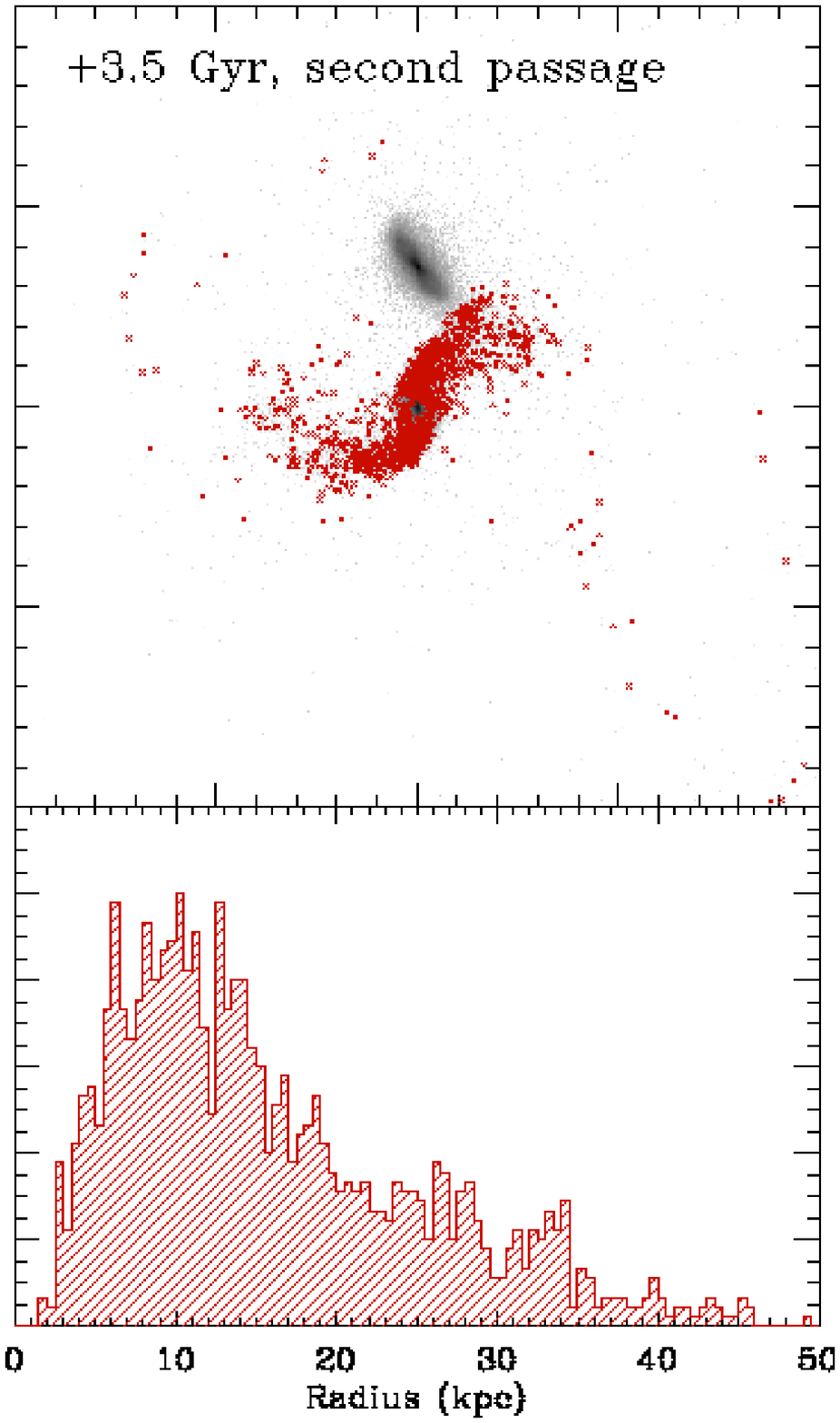}}
\resizebox{4.5cm}{!}{\includegraphics{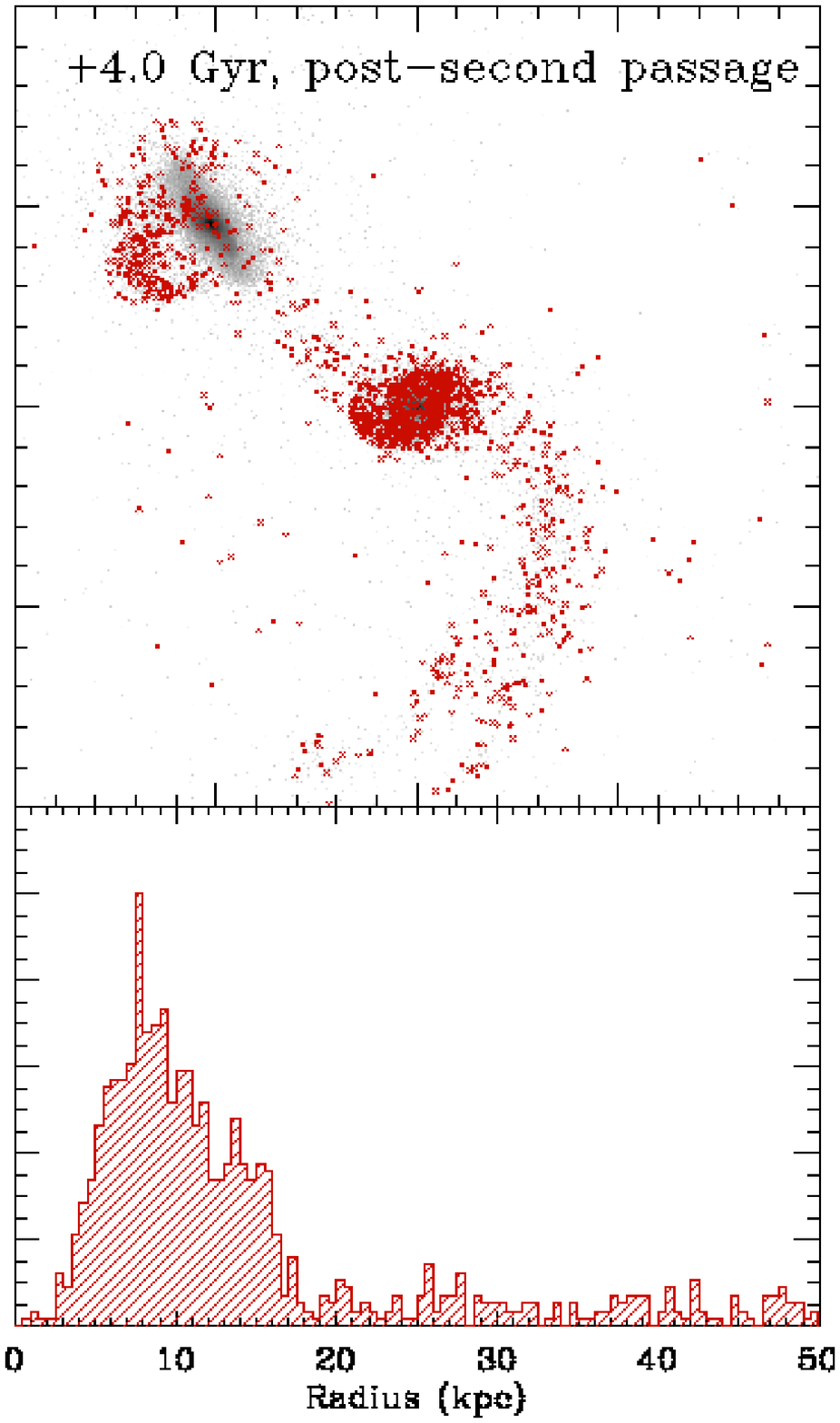}}
\resizebox{4.5cm}{!}{\includegraphics{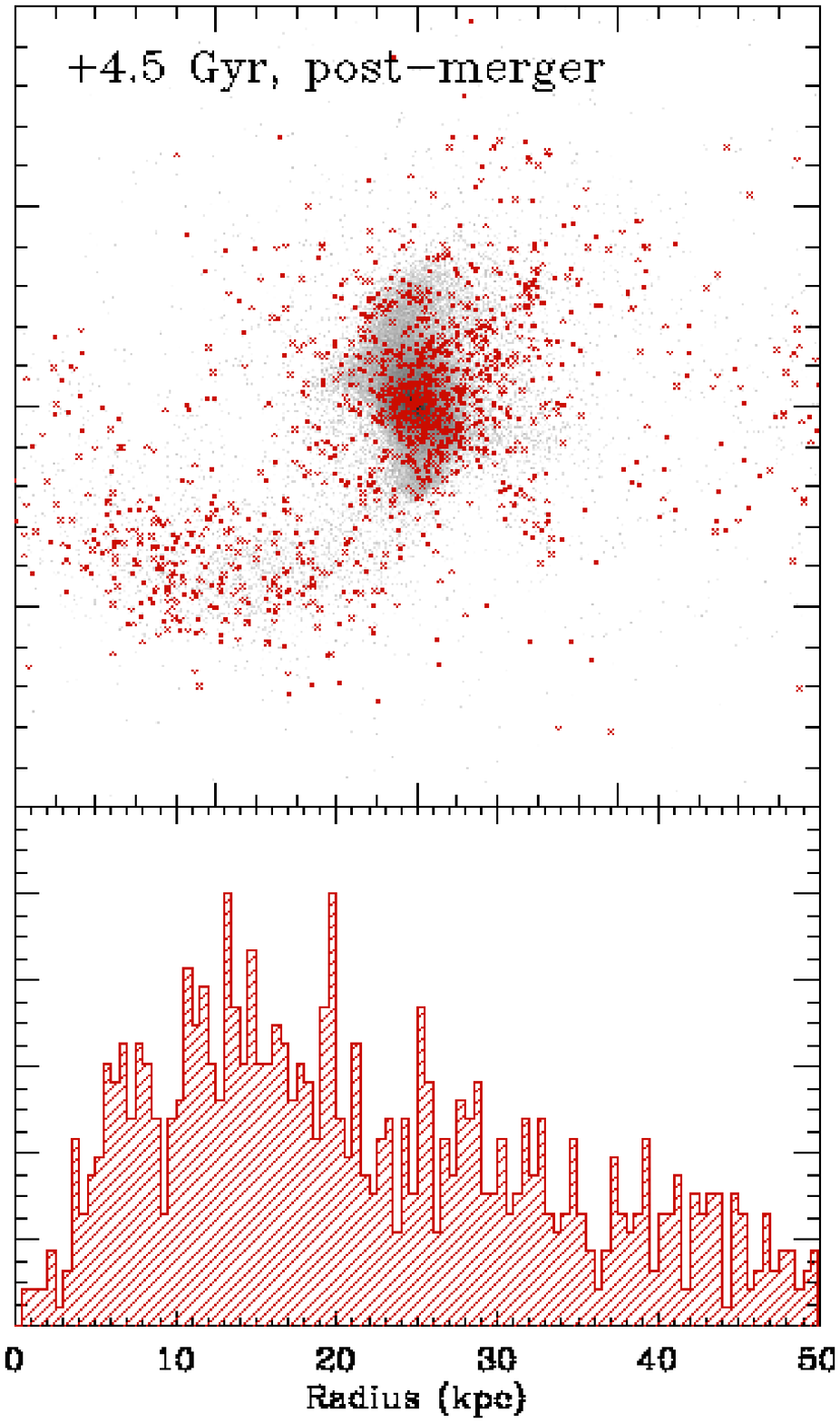}}\\
\caption{ The possible location of the Sun during various stages of the
merger between the Milky Way and Andromeda.  The top panel shows all
stellar particles in our simulation that have a present-day galactocentric
radius of 8$\pm0.1$~kpc with a red cross and tracks their position into
the future.  The bottom panel presents a histogram of the radial distance
from the center of the Milky Way.  }
\label{fig:whe}
\end{center}
\end{figure*}

An interesting consequence of the short timescale for the merger between
the Milky Way and Andromeda is the possibility that a human (or decedent
thereof) observer will witness the interaction and merger.  At the heart of
this issue is the comparison between the lifetime of the Sun and the
interaction timescale.

Current evolutionary models \citep[see, e.g.,][]{SBK93} predict that the
Sun will steadily increase its size and luminosity for the next 7~Gyr as it
slowly consumes all available hydrogen and evolves towards a red giant
phase.  While this places a strong upper limit to the extent of life on
Earth, it is likely that much smaller changes ($<50$\%) in the Sun's
luminosity will significantly alter the Earth's atmosphere and thus its
habitability within the next 1.1-3.5~Gyr \citep{Kas88}.  \citet{Kory01}
suggested that the onset of these effects could be delayed by increasing
the orbital radius of the Earth through a sequence of interactions with
bodies in the outer Solar System, and we can not rule out the possible
colonization of habitable planets in nearby stars, especially long-lived
M-dwarfs \citep{Udry07} whose lifetime may exceed $10^{12}$ years
\citep{ABL05}.  In short, it is conceivable that life may exists for as
little as 1.1~Gyr into the future or, if interstellar travel is possible,
much longer.

Regardless of the prospects for life in the future, we can attempt to
predict what any potential observer at the solar Galactic circle might see
by tracking candidate Suns in our simulation.  In particular, we flag all
stellar particles with a galactocentric orbital radius of 8$\pm0.5$~kpc,
corresponding to the observed value for the Sun \citep{Eis03}, and
subsequently follow the location of these particles forward in time.  This
procedure typically yields $\sim700$ particles as ``candidate Suns''
(although it should be kept in mind that owing to our limited resolution,
each of the simulated particles is many orders of magnitude more massive
than the Sun).  The results are presented in Figure~\ref{fig:whe} and
demonstrate some of the possible outcomes for the future location of our
Sun.  Given the uncertainties in our model parameters and the fact that the
orbital period of the Sun around the Milky Way is much shorter than the
merger timescale, it is not possible to forecast reliably the actual phase
of the Galactic orbit of the Sun at the time of closest approach to
Andromeda.  Therefore we regard all the stellar particles at the
galactocentric radius of the Sun as equally probable of representing the
Sun.  We will first outline some of the general features of our fiducial
model, before showing similar distributions from a subset of models to
assess the reliability of these results.

Figure~\ref{fig:whe} outlines the wide variety of potential locations for
our candidate Suns during the future evolution of the Local Group.  For
example, the top--right panel in Figure~\ref{fig:whe} demonstrates the
location of the candidate Suns after the first passage of Andromeda.  At
this point, the observer will most likely still be in the (now disturbed)
disk of the Milky Way, but there exists a 12\% chance that the Sun will be
tidally ejected and take part in the tidal tail material that is
$>20$~kpc away from the Milky Way center.

The probability that the candidate Sun will be located farther than
$20$~kpc from the Milky Way center steadily increases as the interaction
progresses.  At the second passage, the percentage of Suns that are farther
than $20$~kpc is 30\%.  This number increases to 48\% after the second
passage. and is 68\% in the merger remnant.  In fact, there is a 54\%
chance that the Sun will be at radii larger than 30~kpc in the merger
remnant. However, we caution that this probability is derived at one point
in time.  Individual stars will generally spend much of their orbital time
at large radii, even if their orbit is eccentric and they come much closer
to the galactic center at other times.

One unexpected possibility for the location of the future Sun is
demonstrated in the lower--middle panel of Figure~\ref{fig:whe}, namely the
Sun may actually become bound to Andromeda instead of the Milky Way before
the two galaxies coalesce.  Such a situation occurs when material becomes
loosely bound after the first passage and is later captured by the
gravitational potential of Andromeda during its second passage.  While this
outcome is unexpected and certainly exciting, only 2.7\% of the candidate
Suns became bound to Andromeda and so this outcome is relatively unlikely.

\begin{figure*}
\begin{center}
\resizebox{5.3cm}{!}{\includegraphics{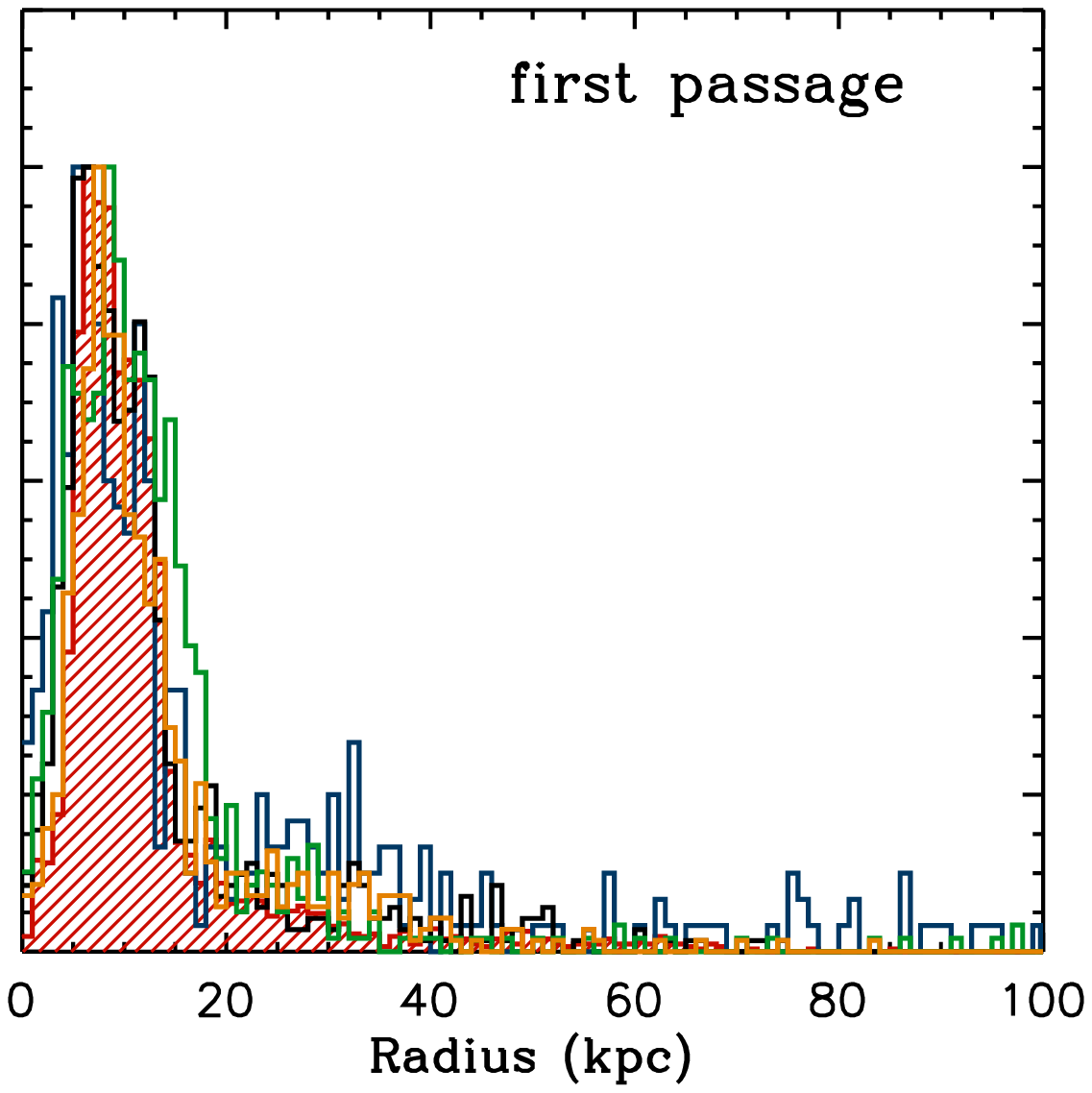}}
\resizebox{5.3cm}{!}{\includegraphics{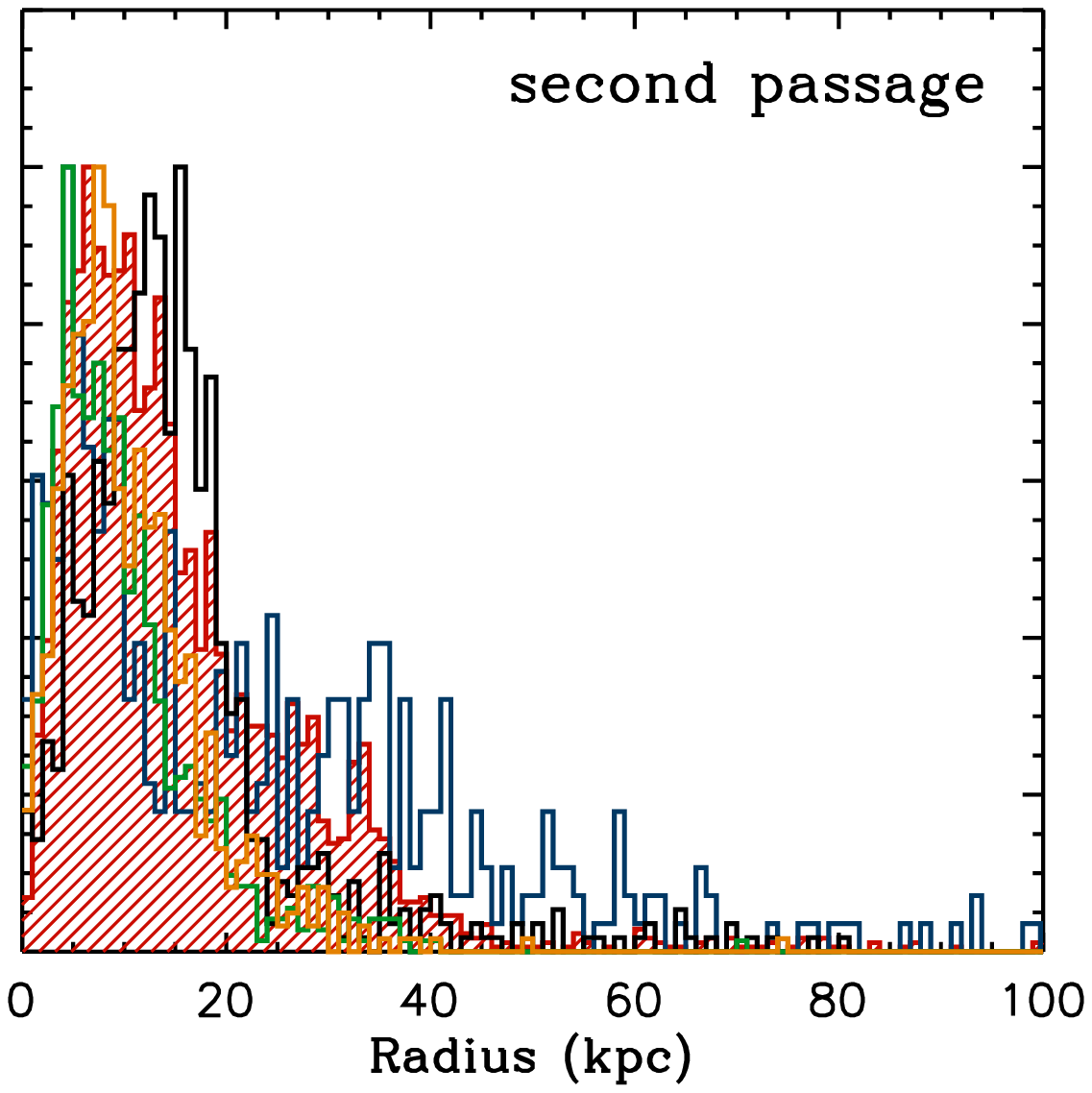}}
\resizebox{5.3cm}{!}{\includegraphics{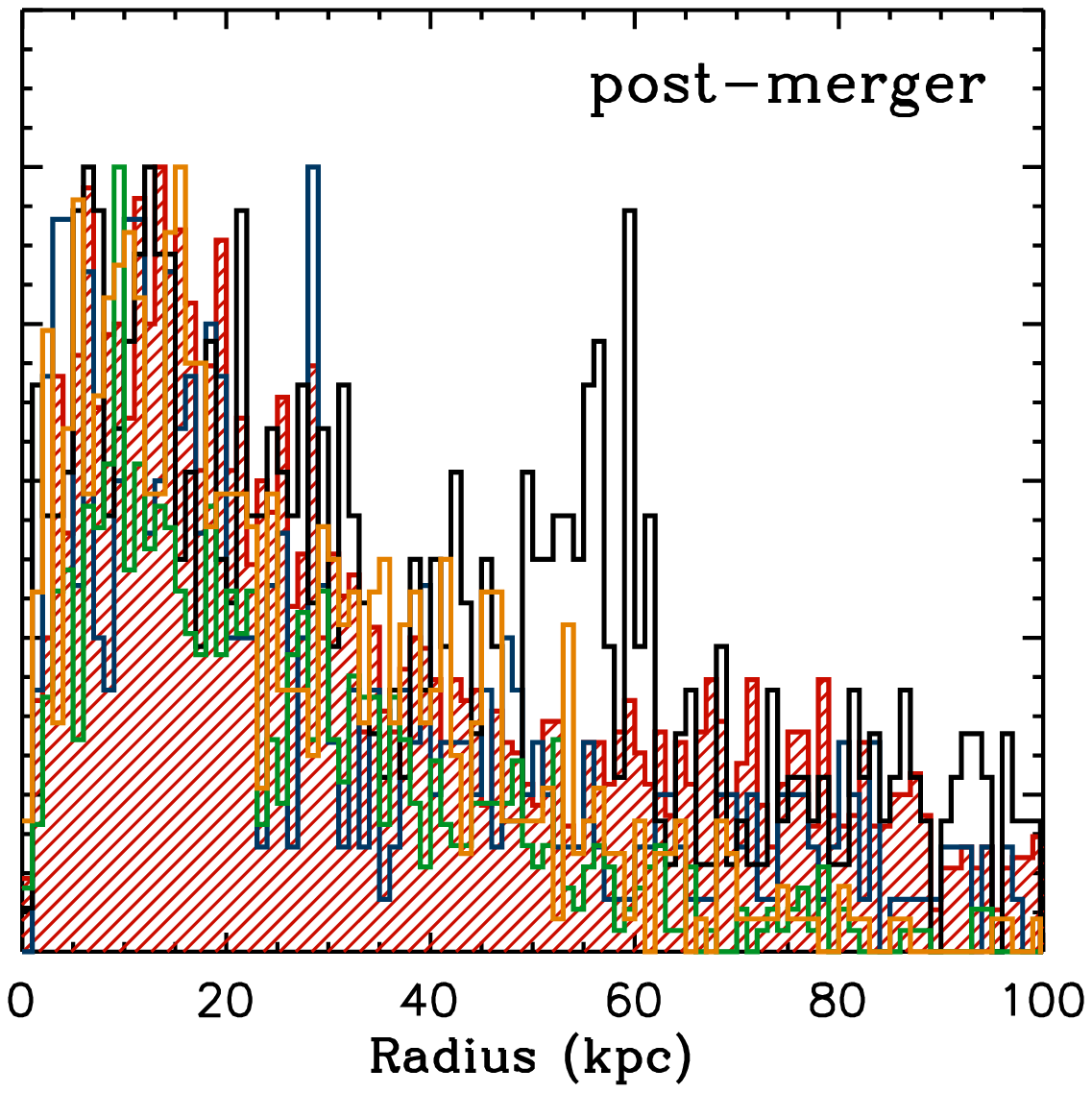}}\\
\caption{ The possible location of the Sun during the first passage (left
panel), second passage (middle panel), and final merger (right panel) for
several different models of the merger between the Milky Way and Andromeda.
The histograms are difficult to distinguish, which highlights the generic
probability for the Sun to reside at various galactocentric radii.  The
distribution of radii from the fiducial model, which is presented in
Figure~\ref{fig:whe}, is denoted by a hatched histogram. }
\label{fig:whe_ensemble}
\end{center}
\end{figure*}

In Figure~\ref{fig:whe_ensemble} we attempt to assess whether the
percentages just quoted for our fiducial model are characteristic of all
our models and are therefore robust.  Each panel in
Figure~\ref{fig:whe_ensemble} shows the distribution of galactocentric
radii for candidate Suns during three periods of the interaction in several
representative Local Group models, including the fiducial one.  The
similarity between the models highlights the generic features of the
fiducial model, although there is a large spread in the percentages quoted
in the previous paragraphs.  The common distributions might have been
expected given the similar orbital evolution demonstrated in
Figure~\ref{fig:csep_ensemble} and the tendency for mergers to preserve the
hierarchy of the initial binding energies of the collisionless particles in
the merger remnant.

We note that several of the interactions generate unique tidal features
which result in subtle differences in the distributions shown in
Figure~\ref{fig:whe_ensemble}.  These differences are an outcome of
different disk-spin orientations that were examined in the different
models.  One model in particular yields a remnant where some of the
candidate Suns reside in a shell--like structure at $\sim50-60$~kpc (see
right--most panel in Figure~\ref{fig:whe_ensemble}).  Only half of the
models include the possibility for the Sun to be ``stolen'' by Andromeda,
however one model yielded a probability that was nearly three times larger
than in the fiducial case.

\subsection{Star Formation}
\label{ssec:sf}

\begin{figure}
\begin{center}
\resizebox{8.0cm}{!}{\includegraphics{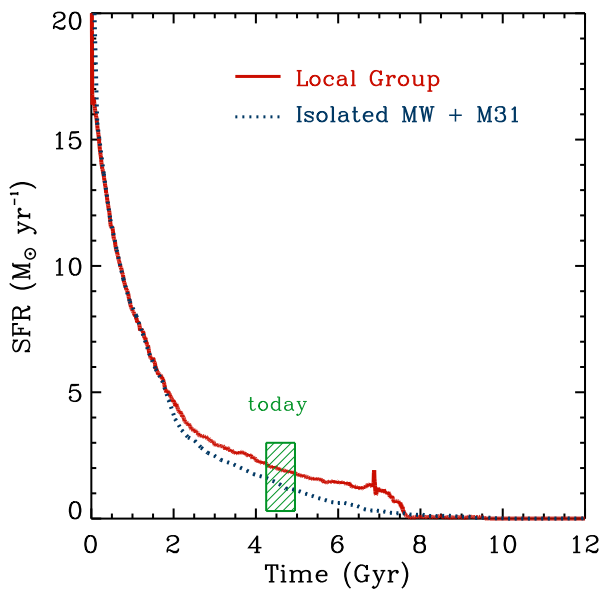}}\\
\caption{
The cumulative star-formation rate during the merger
of the Milky Way and Andromeda compared to the star
formation for models of the Milky Way and Andromeda
evolved in isolation.
}
\label{fig:sfr}
\end{center}
\end{figure}

There is mounting evidence that galaxy interactions are the predominant
mechanism for producing large bursts of star formation, such as in the
ultra--luminous infrared galaxies \citep[ULIRGs,][]{SM96,BH92rev}.  During
the interactions, gravitational torques extract angular momentum from the
gas and funnel it to the center of the merger galaxy where it participates
in a centrally--concentrated starburst.  Because the future of the Local
Group entails a major galaxy merger, it is natural to examine whether the
merger of the Milky Way with Andromeda will become a ULIRG.

This question is addressed explicitly in Figure~\ref{fig:sfr}, which shows
the star--formation rate during the entire evolution of the Local Group,
including the merger between the Milky Way and Andromeda.  Throughout the
entire evolution of the Local Group, the star--formation rate steadily
decreases with time, and hence we conclude that the merger Local Group will
not become a ULIRG in the future.  In fact, the final coalescence yields
star formation that is barely enhanced above the that which would occur if
the Milky Way and Andromeda has not participated in the merger at all.

The weak starburst event triggered by the merger between the Milky Way
and Andromeda is a direct result of their present low gas content.
Moreover, a large fraction ($>75$\%) of this gas will be consumed by
quiescent star formation by the time the merger actually occurs.  In short,
both disks will be extremely gas--poor during the final coalescence and
there will be no fuel for the starburst.

While we have not explicitly tracked the black holes at the center of the
Milky Way and Andromeda, it is interesting to speculate whether the merger
will produce a luminous quasar, which many models argue is intricately
linked to galaxy mergers and starbursts \citep{Hop06big}.  Even though
Figure~\ref{fig:sfr} demonstrates that there is not enough gas to fuel a
powerful starburst, this gas content is clearly sufficient to ignite a
luminous quasar if $\sim1$\% of it is accreted by the black hole.  While
the current work cannot address this possibility in detail, our model
provides a framework to study the formation of quasars in the future.

\subsection{The Merger Remnant}
\label{ssec:mrem}

\begin{figure*}
\begin{center}
\resizebox{6.0cm}{!}{\includegraphics{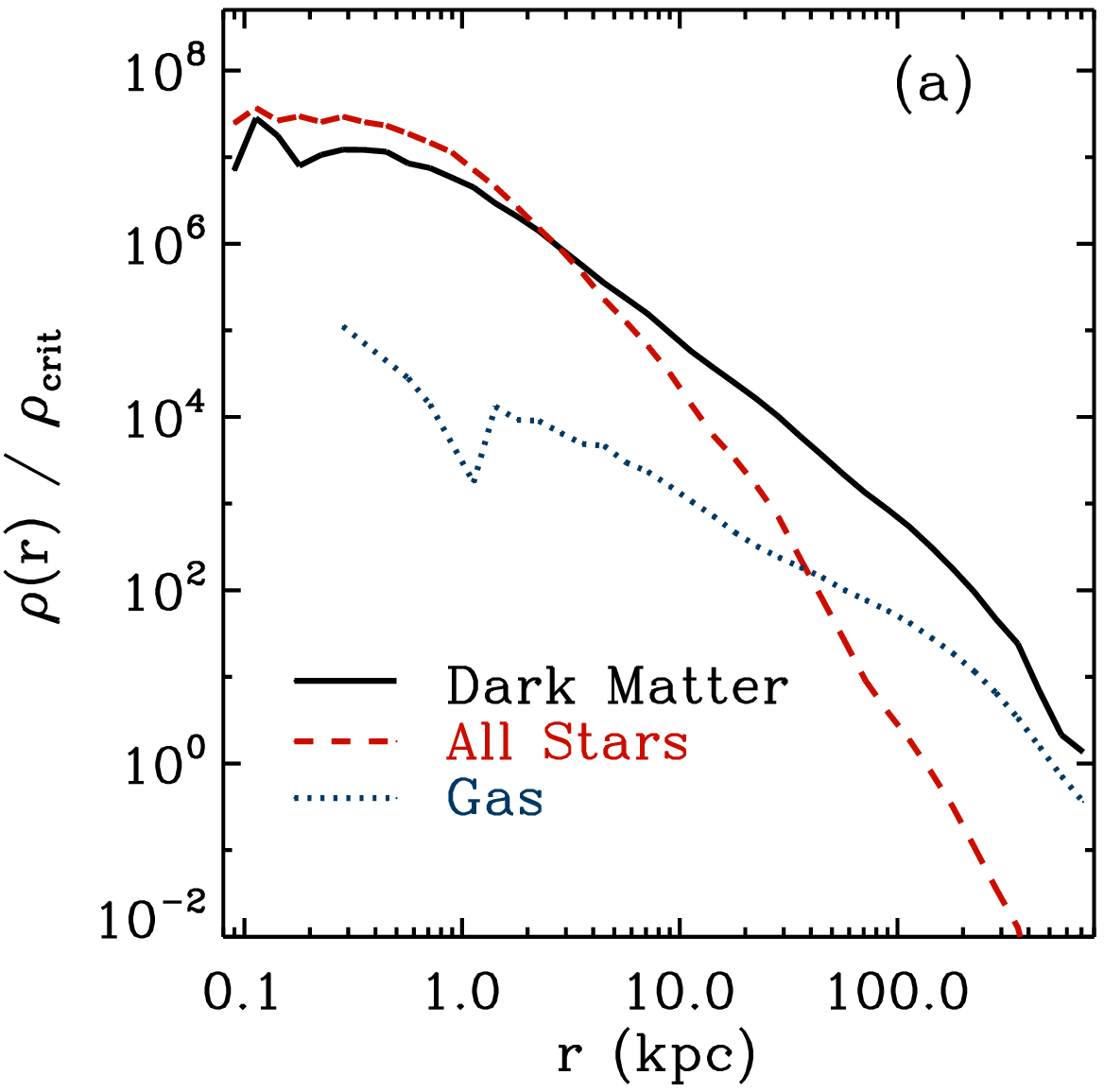}}
\resizebox{6.0cm}{!}{\includegraphics{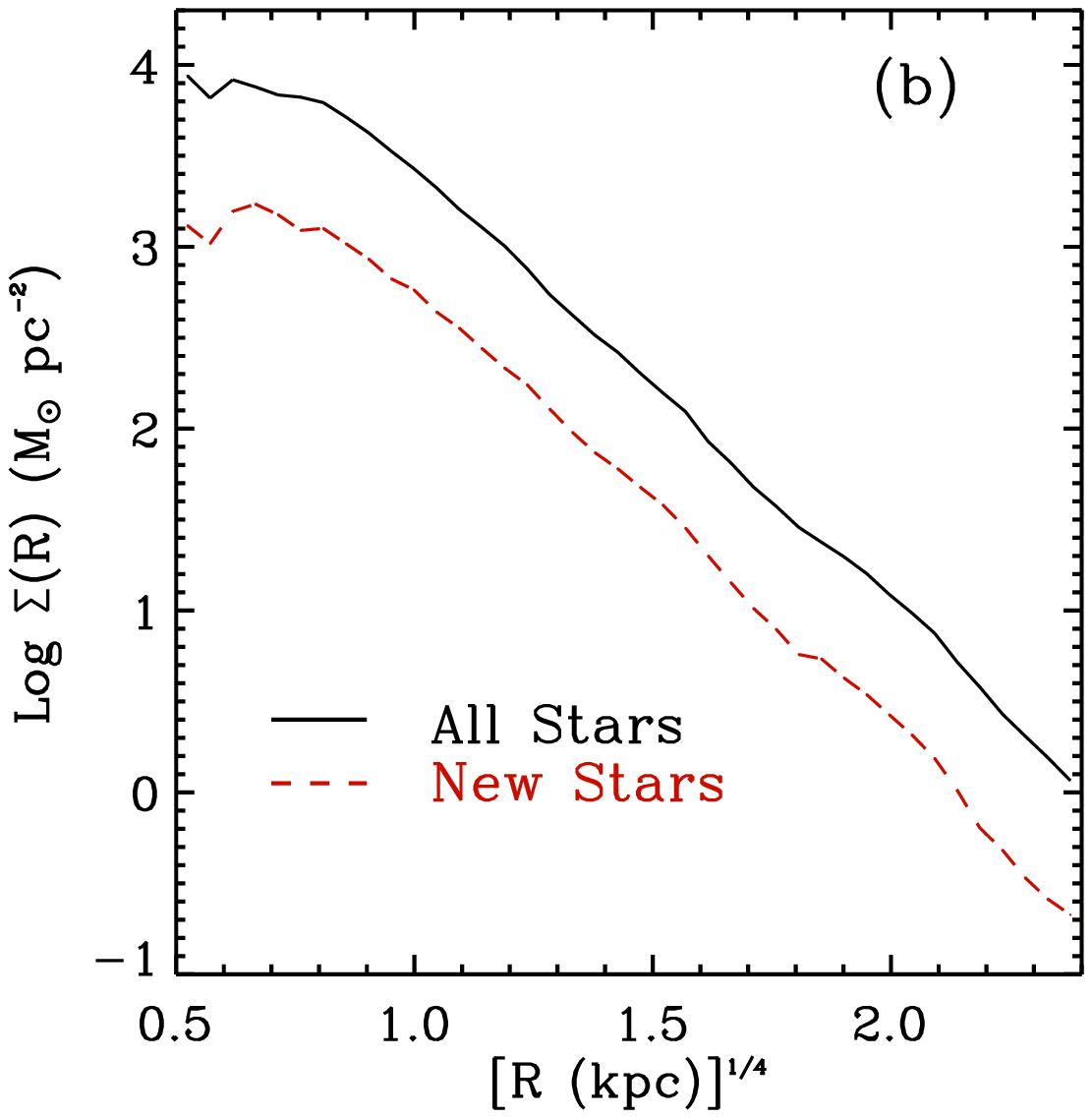}}\\
\resizebox{6.0cm}{!}{\includegraphics{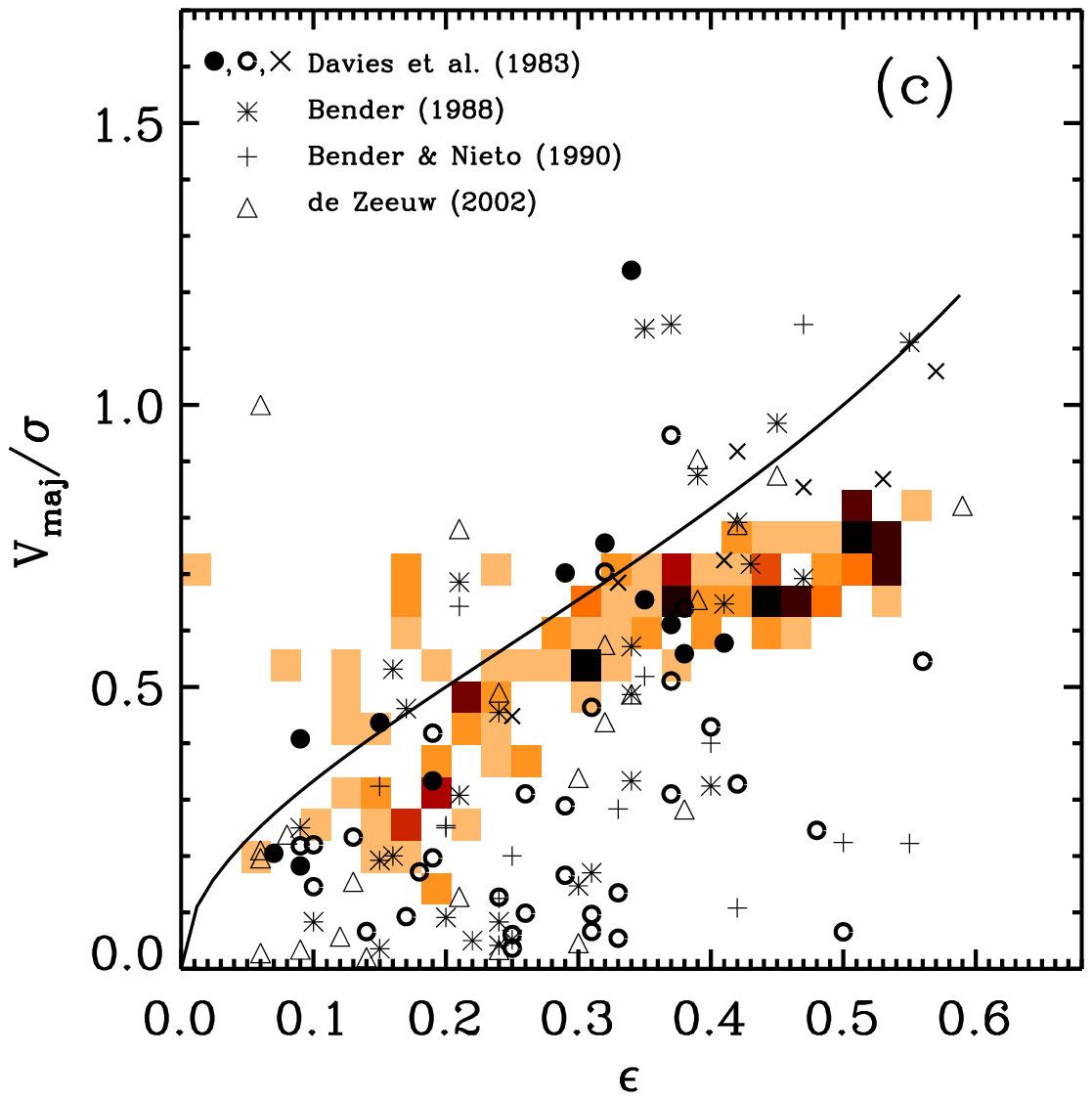}}
\resizebox{6.0cm}{!}{\includegraphics{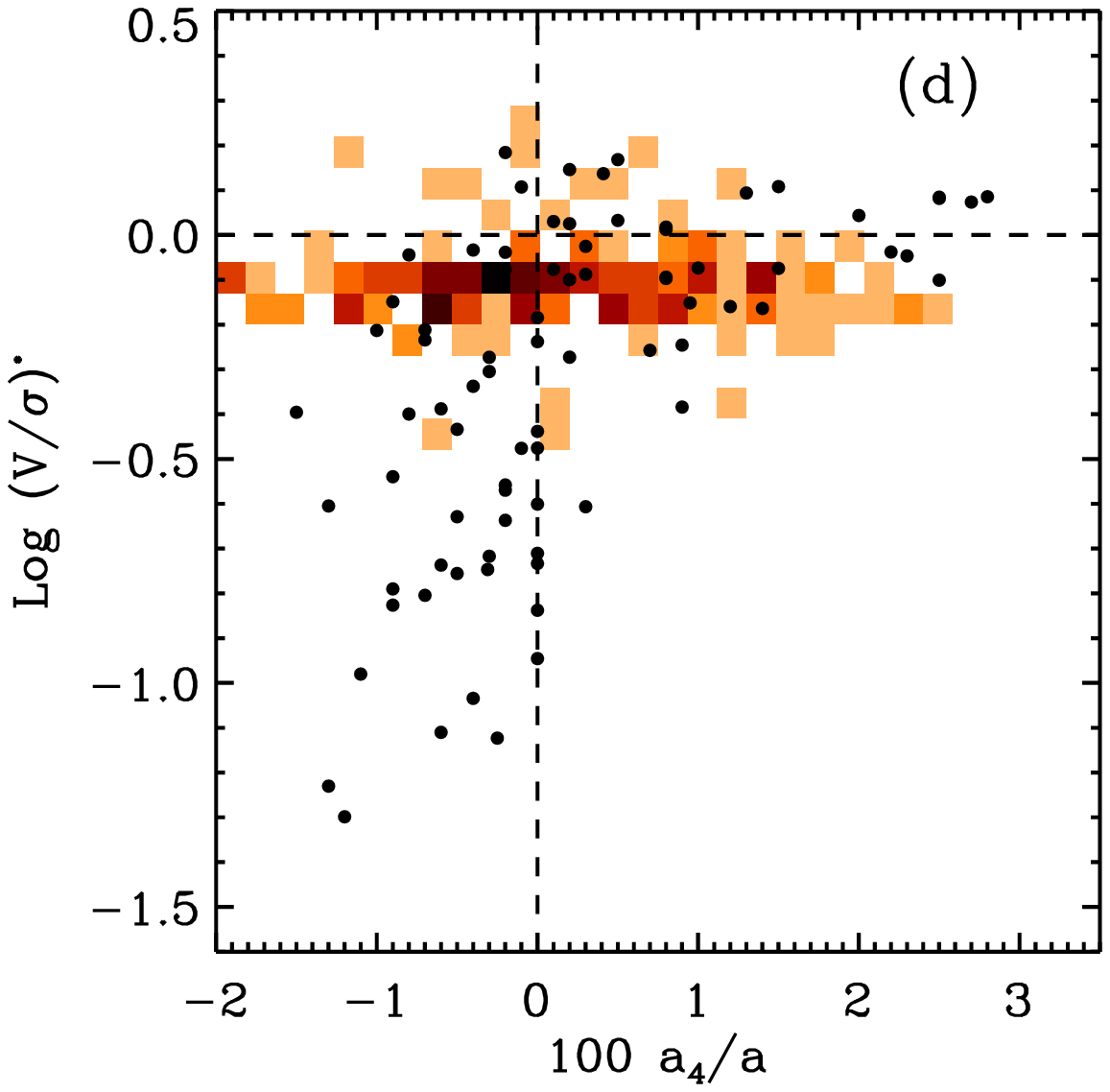}}\\
\resizebox{6.0cm}{!}{\includegraphics{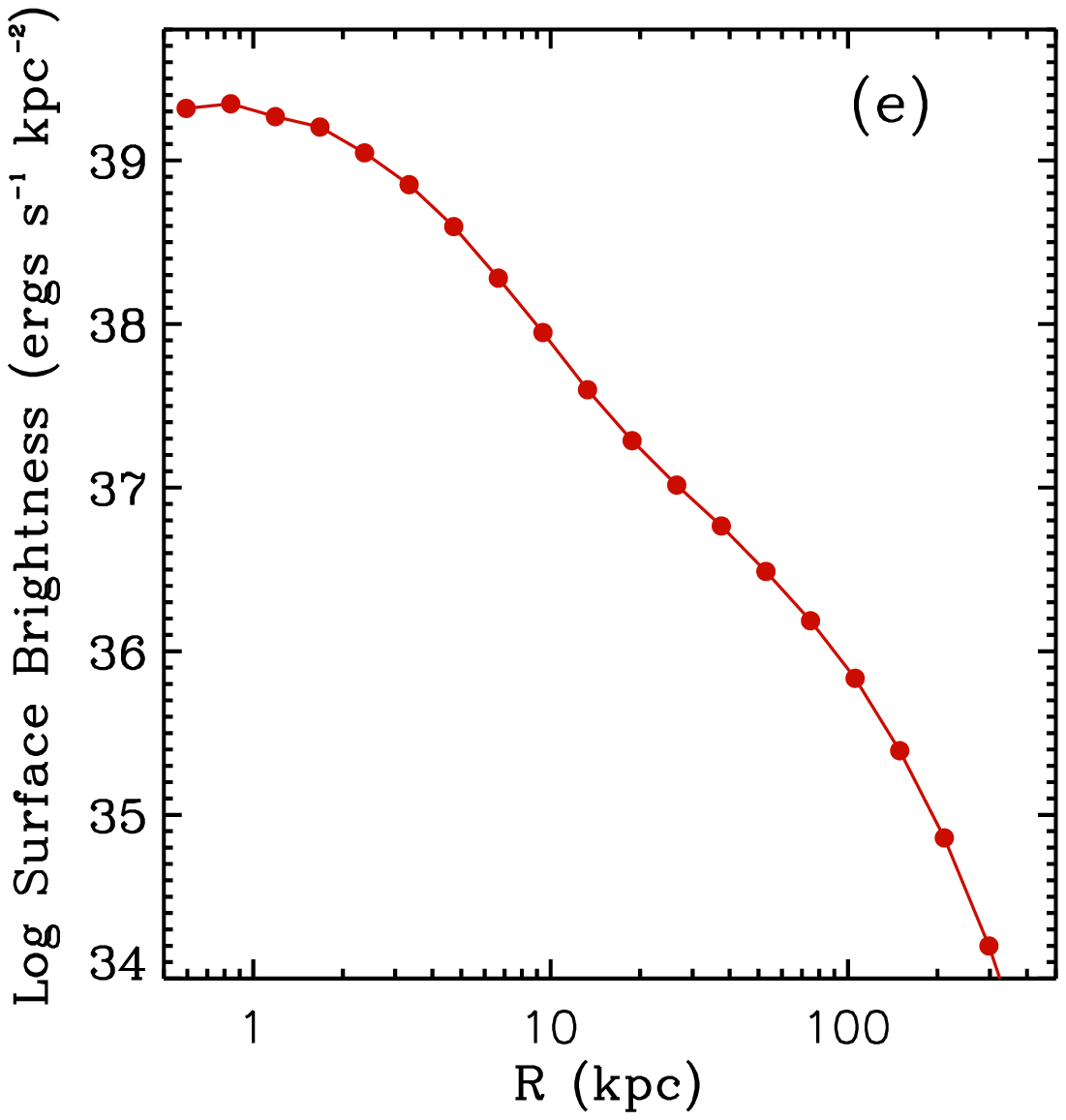}}
\resizebox{6.0cm}{!}{\includegraphics{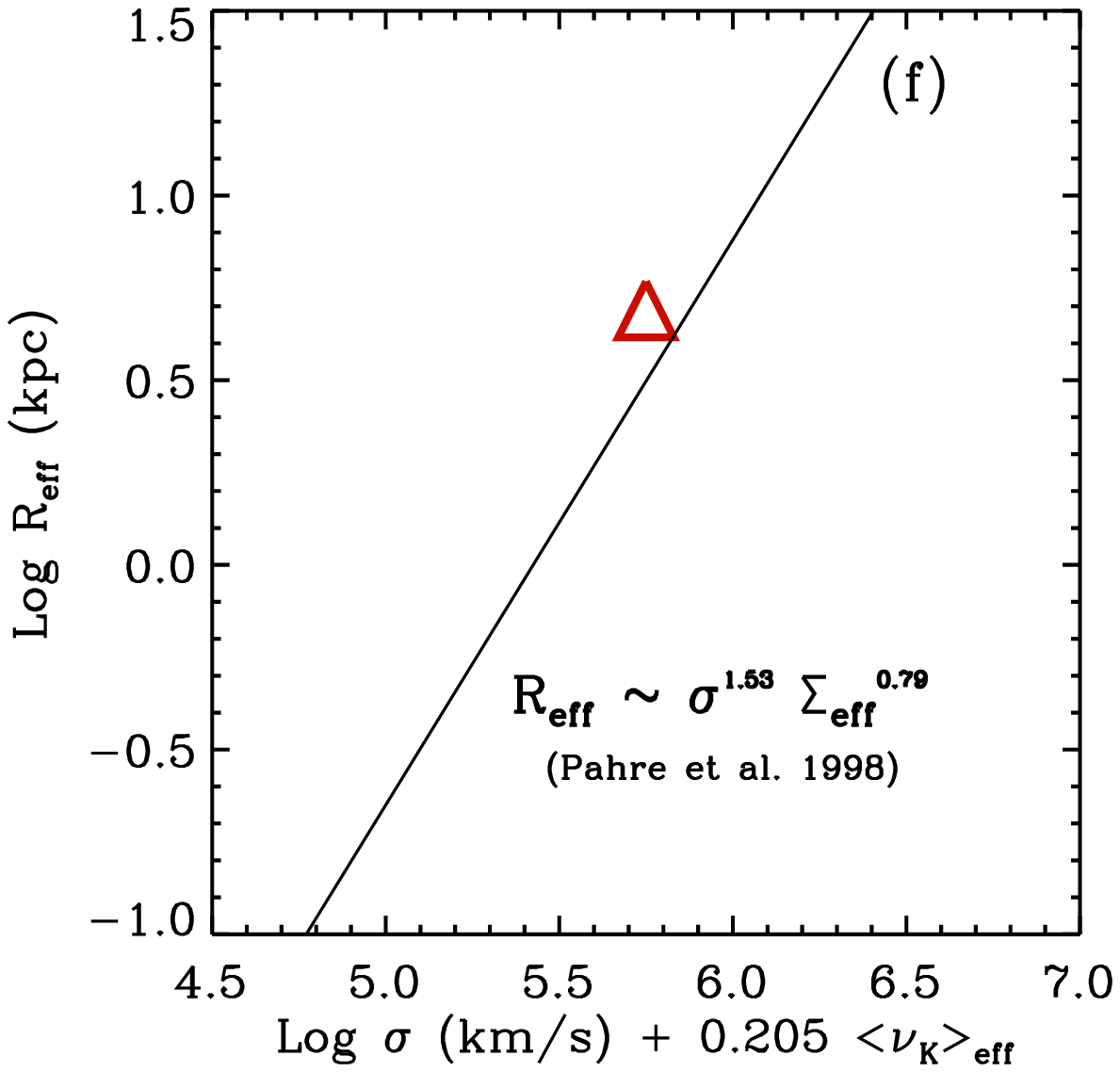}}\\
\caption{ Properties of the Milky Way and Andromeda merger remnant, which
we abbreviate as {\it Milkomeda}.  Panel (a) shows the spherically averaged
distribution of dark matter, stars, and gas.  The mass density is expressed
in units of the present--day critical density, $\rho_{\rm crit}=
10^{-29}~{\rm g~cm^{-3}}$.  Panel (b) shows the projected distribution of
stellar mass, including the contribution from stars formed during the
course of the simulation which are termed ``New Stars,'' plotted against
$R^{1/4}$, where $R$ is the projected radius.  Panel (c) presents the
kinematics of {\it Milkomeda}.  For 195 projections, the ellipticity of the
half--mass isophote $\varepsilon$, the rotation velocity ($V$) along the major
axis $V_{\rm maj}$ and the central velocity dispersion $\sigma$, are plotted
as a colored histogram, with darker shades representing a higher concentration
of projected properties.  The solid line represents what is expected from an
oblate isotropic rotation and overplotted is data extracted from the literature.
Panel (d) shows the velocity anisotropy, defined as the ratio $V/\sigma$
normalized by the value expected for an oblate isotropic rotator, versus the
isophotal shape parameter $a_4/a$, where positive values are disky isophotes and
negative values are boxy.  Overplotted in black are current data (Ralf Bender,
private communication).  Panel (e) shows the X-ray surface brightness profile,
and panel (f) shows {\it Milkomeda} on the near--IR fundamental plane along with
the relation observed by \citet{Pah98}.  }
\label{fig:remnant}
\end{center}
\end{figure*}

Galaxy mergers have become an area of intense study owing to their proposed
role in shaping galaxy morphology.  In particular, the ``merger
hypothesis'' \citep{T77} posits that the interaction and merger of two
spiral galaxies leaves behind a remnant that is morphologically and
kinematically similar to an elliptical galaxy.  Since the future evolution
of the Local Group contains such an event, it is natural to examine
whether the local group will eventually consist of a single elliptical
galaxy, and if so, how do the properties of this galaxy differ from 
present--day ellipticals which were formed at earlier epochs.

Note that the morphology of the Milky Way and Andromeda merger remnant,
which we abbreviate hereafter as {\it Milkomeda}, was presented in
Figure~\ref{fig:zoomimage}.  This figure confirms the notion that galaxy
mergers leave behind remnants that are spheroidal in shape, contain stars
with a wide range of orbits and a large velocity dispersion, and possess very
modest net rotation.

In Figure~\ref{fig:remnant}, we present a series of plots that further
support the assertion that {\it Milkomeda} resembles an elliptical galaxy
and also serve to better quantify its properties.  In panel (a), we present
the spherically averaged mass profile, decomposed by the three components;
dark matter, stellar and gaseous mass.  At radii greater than $\sim2-3$~kpc,
the dark matter dominates the mass density, and at radii greater than
$\sim20$~kpc, the profile is well--fit by the NFW \citep{NFW96} or Hernquist
\citep{H90} profile.

The projected mass distribution shown in panel (b) of 
Figure~\ref{fig:remnant} presents the first direct evidence that 
{\it Milkomeda} resembles an elliptical galaxy.  Specifically, the stellar
surface density is close to a pure R$^{1/4}$ distribution, with the exception
of the inner $\sim400$~pc ($\sim2.5\times$ the gravitational softening
length), where the surface density flattens to a nearly constant
density core.

We have also quantified the kinematics and isophotal shape of the {\it
Milkomeda} galaxy following the methods of \citet{Cox06rot}.  Panel (c) in
Figure~\ref{fig:remnant} presents the anisotropy diagram, a measure of the
half--mass isophote ellipticity versus the maximum rotation along its major
axis divided by the central velocity dispersion.  The shaded region
represents the distribution of values for {\it Milkomeda} if viewed
from 195 directions that uniformly sample the unit sphere \citep[with
angles selected using HEALPIX,][]{Gor05}.  Also plotted in this figure is
the relation expected from an oblate isotropic rotator as a solid line.
Owing to the high concentration of projections that closely track the solid
line, this analysis demonstrates that {\it Milkomeda} is nearly an oblate
isotropic rotator.  In addition, the
deviations from a perfect ellipse are also quantified and are presented in
panel (d) in Figure~\ref{fig:remnant}.  Depending on the viewing direction,
{\it Milkomeda} may appear to be either disky or boxy, with a slightly larger
fraction of views producing disky isophotes.

We have also analyzed the properties of the hot gas in and around {\it
Milkomeda}.  The evolution of this component was presented in
Figure~\ref{fig:gasimages} and clearly demonstrates the formation of an
extended gaseous halo, primarily accreted from the large reservoir of the
intragroup medium.  Although the gas temperature was originally
$3\times10^5$~K, it has been shock-heated to the virial temperature of
$\sim3\times10^6$~K in {\it Milkomeda} and has a gradient to cooler
temperatures at large radii.  This hot gas leads to an X-ray surface
brightness profile shown in panel (e) of Figure~\ref{fig:remnant} and
a total X-ray luminosity of
$\sim10^{41}$~ergs~s$^{-1}$, which is consistent with present--day
elliptical galaxies of equivalent B-band luminosity 
($\sim3\times10^{10}$~\lsun).

In general, {\it Milkomeda} resembles the remnants of gas--rich major
mergers, which in--turn resemble the general population of low-- and
moderate--luminosity elliptical galaxies \citep{NJB06,Cox06rot}.  However,
there are some systematic differences that likely arise because of the much
smaller gas content of the Milky Way and Andromeda when they merge.  In
particular, the inner regions of {\it Milkomeda} have a much lower stellar
density than present--day ellipticals, which are often observed to have
steep spikes of newly formed stars \citep{RJ04,Kor07}.  These central
excesses arise from high gas concentrations that fuel nuclear starbursts
during the merger event \citep[see,e.g.,][]{MH94dsc,Sp00,Cox06}. This
process does not occur during the formation of {\it Milkomeda} (see
\S\ref{ssec:sf}), for which the stars formed during the merger simulation
have an identical profile to the entire stellar population (see panel (b)
in Figure~\ref{fig:remnant}).

The diffuse nature of {\it Milkomeda} is also evident if it is projected
onto the near-IR fundamental plane as shown in panel (f) of 
Figure~\ref{fig:remnant}, where it lies above the observed relation.  The
half--mass radius is 4.9~kpc, which is larger than the mean relation found in
the Sloan Digital Sky Survey \citep{Shen03,Des07} for galaxies of
equivalent stellar mass ($1.3 \times 10^{12}$\msun) or r--band absolute
magnitude (-21.2).  These comparisons gives credence to the claim that
present--day ellipticals can not have formed from the merger of present day
spirals \citep{NO07}.

\section{Conclusions}
\label{sec:conc}

In this paper we have used an N-body/hydrodynamic simulation to track the
evolution of the Local Group, focusing primarily upon the two most massive
galaxies: the Milky Way and Andromeda.  In contrast to most prior work,
which typically employed models for the Local Group to infer its total mass
or the proper motion of its constituents, we simulated the large-scale
dynamics of all matter in the Local Group including dynamical friction on
the intragroup medium, leading to the eventual merger of the Milky Way and
Andromeda.

Owing to the diffuse intragroup medium that was assumed to pervade the
Local Group with a total mass comparable to that the galaxies, we have
found that the interaction and merger between the Milky Way and Andromeda
will occur in less than 5~Gyr, a timescale comparable to the lifetime of
the Sun \citep{SBK93}.  This Local Group model therefore admits
the possibility that future astronomers in the Solar System will witness
parts of, or the entire interaction and merger of the Milky Way and
Andromeda.

With this in mind we have calculated the probable location of our Solar
System during specific points of the future interaction and find several
interesting outcomes.  First, there is a chance that the Sun will be
ejected along with other tidal material into a long tidal tail following
the first passage of Andromeda.  Second, as a result of the disruptive
effects of each close tidal passage, there is an increasing chance that the
Sun will inhabit extended tidal features as the interaction proceeds.
Moreover, there is a small chance that the Sun will be more tightly bound
to Andromeda at some point during the merger.  In such a case, Andromeda
will capture the Sun and future astronomers in the solar system might see
the Milky Way as an external galaxy in the night sky\footnote{Aside from
the changed night sky during the interaction between the Milky Way and
Andromeda, future observers might witness enhanced comet showers due to the
increased flux of stars passing by the solar system and perturbing the Oort
cloud.  Additional observables would include a slightly enhanced star
formation in the two galaxies and the production of hot gas by the shocks
surrounding the interaction region.}.

While this paper highlights the possible outcomes of the future interaction
between the Milky Way and Andromeda, we emphasize that our model is likely
to be one of many plausible models within an ensemble of possibilities that
span the uncertain value of the transverse velocity of Andromeda and the
density (and other properties) of the intragroup medium.  We have performed
20 additional runs in order to test the sensitivity of our results to
various assumptions of our model -- mainly involving the initial orbit of
the Milky Way and Andromeda.  These runs yield similar estimates for the
merger timescale as well as for the possible locations of the Sun in the
future, provided that the intragroup medium is indeed similar to our
fiducial case.  While this gives us some confidence that our results are
robust, an even larger suite of models, that spans a much wider set of
model assumptions, will provide better statistics on these results. In
addition, employing higher resolution simulations with increased
complexity, may shed light on the nature of the intragroup medium, the soft
X-ray background \citep{Oso02}, galactic substructure \citep{Will05}, the
origin of the Magellanic Clouds and Stream \citep{Besla07}, and the future
evolution of globular clusters \citep{For00}.

Finally, we note that the simulated views from the distribution of
locations for the candidate Suns in the merger remnant (see
Fig.~\ref{fig:whe}), which we have termed {\it Milkomeda}, represent
the {\it only} views available for a future
local astronomer. Extragalactic astronomy will come to an end within 100
billion years if the cosmological constant will not evolve with time. Owing
to the accelerated expansion caused by a steady cosmological constant, all
galaxies not bound to the Local Group will eventually recede away from the
Local Group and exit our event horizon \citep{Loeb02}.  At that point, the
merger product of the Milky Way and Andromeda (with its bound satellites)
will constitute the entire visible Universe \citep{NL03}.

\section*{Acknowledgments}

The simulations were performed at the Center for Parallel Astrophysical
Computing at the Institute for Theory and Computation at the
Harvard-Smithsonian Center for Astrophysics.  This research was supported
in part by a grant from The Foundational Questions Institute.  We
acknowledge helpful discussions with G. Besla, J. Dubinski, S. Dutta, 
L. Hernquist, P. Hopkins, and B. Robertson.  We thank Ralf Bender for
kindly providing data used in Figure~\ref{fig:remnant}.

\bibliographystyle{mn2e}
\bibliography{lg}

\begin{thebibliography}{}

\bibitem[\protect\citeauthoryear{{Adams}, {Bodenheimer} \& {Laughlin}}{{Adams}
  et~al.}{2005}]{ABL05}
{Adams} F.~C.,  {Bodenheimer} P.,    {Laughlin} G.,  2005, Astronomische
  Nachrichten, 326, 913

\bibitem[\protect\citeauthoryear{{Barnes}}{{Barnes}}{1988}]{B88}
{Barnes} J.~E.,  1988, \apj, 331, 699

\bibitem[\protect\citeauthoryear{{Barnes} \& {Hernquist}}{{Barnes} \&
  {Hernquist}}{1992}]{BH92rev}
{Barnes} J.~E.,  {Hernquist} L.,  1992, \araa, 30, 705

\bibitem[\protect\citeauthoryear{{Barnes} \& {Hernquist}}{{Barnes} \&
  {Hernquist}}{1991}]{BH91}
{Barnes} J.~E.,  {Hernquist} L.~E.,  1991, \apjl, 370, L65

\bibitem[\protect\citeauthoryear{{Besla}, {Kallivayalil}, {Hernquist},
  {Robertson}, {Cox}, {van der Marel} \& {Alcock}}{{Besla}
  et~al.}{2007}]{Besla07}
{Besla} G.,  {Kallivayalil} N.,  {Hernquist} L.,  {Robertson} B.,  {Cox} T.~J.,
   {van der Marel} R.~P.,    {Alcock} C.,  2007, ArXiv Astrophysics e-prints

\bibitem[\protect\citeauthoryear{{Binney} \& {Tremaine}}{{Binney} \&
  {Tremaine}}{1987}]{BT}
{Binney} J.,  {Tremaine} S.,  1987, {Galactic dynamics}.
Princeton, NJ, Princeton University Press, 1987, 747 p.

\bibitem[\protect\citeauthoryear{{Brunthaler}, {Reid}, {Falcke}, {Greenhill} \&
  {Henkel}}{{Brunthaler} et~al.}{2005}]{Bru05}
{Brunthaler} A.,  {Reid} M.~J.,  {Falcke} H.,  {Greenhill} L.~J.,    {Henkel}
  C.,  2005, Science, 307, 1440

\bibitem[\protect\citeauthoryear{{Busha}, {Adams}, {Wechsler} \&
  {Evrard}}{{Busha} et~al.}{2003}]{Bus03}
{Busha} M.~T.,  {Adams} F.~C.,  {Wechsler} R.~H.,    {Evrard} A.~E.,  2003,
  \apj, 596, 713

\bibitem[\protect\citeauthoryear{{Busha}, {Evrard}, {Adams} \&
  {Wechsler}}{{Busha} et~al.}{2005}]{Bus05}
{Busha} M.~T.,  {Evrard} A.~E.,  {Adams} F.~C.,    {Wechsler} R.~H.,  2005,
  \mnras, 363, L11

\bibitem[\protect\citeauthoryear{{Cen} \& {Ostriker}}{{Cen} \&
  {Ostriker}}{1999}]{CO99}
{Cen} R.,  {Ostriker} J.~P.,  1999, \apj, 514, 1

\bibitem[\protect\citeauthoryear{{Cox}, {Dutta}, {Di Matteo}, {Hernquist},
  {Hopkins}, {Robertson} \& {Springel}}{{Cox} et~al.}{2006}]{Cox06rot}
{Cox} T.~J.,  {Dutta} S.,  {Di Matteo} T.,  {Hernquist} L.,  {Hopkins} P.~F.,
  {Robertson} B.,    {Springel} V.,  2006, \apj, 650, 791

\bibitem[\protect\citeauthoryear{{Cox}, {Jonsson}, {Primack} \&
  {Somerville}}{{Cox} et~al.}{2006}]{Cox06}
{Cox} T.~J.,  {Jonsson} P.,  {Primack} J.~R.,    {Somerville} R.~S.,  2006,
  \mnras, 373, 1013

\bibitem[\protect\citeauthoryear{{Dav{\'e}}, {Cen}, {Ostriker}, {Bryan},
  {Hernquist}, {Katz}, {Weinberg}, {Norman} \& {O'Shea}}{{Dav{\'e}}
  et~al.}{2001}]{Dav01}
{Dav{\'e}} R.,  {Cen} R.,  {Ostriker} J.~P.,  {Bryan} G.~L.,  {Hernquist} L.,
  {Katz} N.,  {Weinberg} D.~H.,  {Norman} M.~L.,    {O'Shea} B.,  2001, \apj,
  552, 473

\bibitem[\protect\citeauthoryear{{Desroches}, {Quataert}, {Ma} \&
  {West}}{{Desroches} et~al.}{2007}]{Des07}
{Desroches} L.-B.,  {Quataert} E.,  {Ma} C.-P.,    {West} A.~A.,  2007, \mnras,
  377, 402

\bibitem[\protect\citeauthoryear{{Dubinski}}{{Dubinski}}{2006}]{Dub06}
{Dubinski} J.,  2006, \skytel, 112, 30

\bibitem[\protect\citeauthoryear{{Dubinski}, {Mihos} \& {Hernquist}}{{Dubinski}
  et~al.}{1996}]{Dub96}
{Dubinski} J.,  {Mihos} J.~C.,    {Hernquist} L.,  1996, \apj, 462, 576

\bibitem[\protect\citeauthoryear{{Eisenhauer}, {Sch{\"o}del}, {Genzel}, {Ott},
  {Tecza}, {Abuter}, {Eckart} \& {Alexander}}{{Eisenhauer}
  et~al.}{2003}]{Eis03}
{Eisenhauer} F.,  {Sch{\"o}del} R.,  {Genzel} R.,  {Ott} T.,  {Tecza} M.,
  {Abuter} R.,  {Eckart} A.,    {Alexander} T.,  2003, \apjl, 597, L121

\bibitem[\protect\citeauthoryear{{Fich} \& {Tremaine}}{{Fich} \&
  {Tremaine}}{1991}]{FT91}
{Fich} M.,  {Tremaine} S.,  1991, \araa, 29, 409

\bibitem[\protect\citeauthoryear{{Forbes}, {Masters}, {Minniti} \&
  {Barmby}}{{Forbes} et~al.}{2000}]{For00}
{Forbes} D.~A.,  {Masters} K.~L.,  {Minniti} D.,    {Barmby} P.,  2000, \aap,
  358, 471

\bibitem[\protect\citeauthoryear{{Gao}, {White}, {Jenkins}, {Stoehr} \&
  {Springel}}{{Gao} et~al.}{2004}]{Gao04a}
{Gao} L.,  {White} S.~D.~M.,  {Jenkins} A.,  {Stoehr} F.,    {Springel} V.,
  2004, \mnras, 355, 819

\bibitem[\protect\citeauthoryear{{G{\'o}rski}, {Hivon}, {Banday}, {Wandelt},
  {Hansen}, {Reinecke} \& {Bartelmann}}{{G{\'o}rski} et~al.}{2005}]{Gor05}
{G{\'o}rski} K.~M.,  {Hivon} E.,  {Banday} A.~J.,  {Wandelt} B.~D.,  {Hansen}
  F.~K.,  {Reinecke} M.,    {Bartelmann} M.,  2005, \apj, 622, 759

\bibitem[\protect\citeauthoryear{{Gott} III \& {Thuan}}{{Gott} \&
  {Thuan}}{1978}]{GT78}
{Gott} III J.~R.,  {Thuan} T.~X.,  1978, \apj, 223, 426

\bibitem[\protect\citeauthoryear{{Hellsten}, {Gnedin} \&
  {Miralda-Escud{\'e}}}{{Hellsten} et~al.}{1998}]{HGM98}
{Hellsten} U.,  {Gnedin} N.~Y.,    {Miralda-Escud{\'e}} J.,  1998, \apj, 509,
  56

\bibitem[\protect\citeauthoryear{{Hernquist}}{{Hernquist}}{1990}]{H90}
{Hernquist} L.,  1990, \apj, 356, 359

\bibitem[\protect\citeauthoryear{{Hernquist}}{{Hernquist}}{1993a}]{H93}
{Hernquist} L.,  1993a, \apjs, 86, 389

\bibitem[\protect\citeauthoryear{{Hernquist}}{{Hernquist}}{1993b}]{H93sph}
{Hernquist} L.,  1993b, \apj, 404, 717

\bibitem[\protect\citeauthoryear{{Hopkins}, {Hernquist}, {Cox}, {Di Matteo},
  {Robertson} \& {Springel}}{{Hopkins} et~al.}{2006}]{Hop06big}
{Hopkins} P.~F.,  {Hernquist} L.,  {Cox} T.~J.,  {Di Matteo} T.,  {Robertson}
  B.,    {Springel} V.,  2006, \apjs, 163, 1

\bibitem[\protect\citeauthoryear{{Kahn} \& {Woltjer}}{{Kahn} \&
  {Woltjer}}{1959}]{KW59}
{Kahn} F.~D.,  {Woltjer} L.,  1959, \apj, 130, 705

\bibitem[\protect\citeauthoryear{{Kasting}}{{Kasting}}{1988}]{Kas88}
{Kasting} J.~F.,  1988, Icarus, 74, 472

\bibitem[\protect\citeauthoryear{{Katz}}{{Katz}}{1992}]{Kz92}
{Katz} N.,  1992, \apj, 391, 502

\bibitem[\protect\citeauthoryear{{Kennicutt}}{{Kennicutt}}{1998}]{Kenn98}
{Kennicutt} R.~C.,  1998, \apj, 498, 541

\bibitem[\protect\citeauthoryear{{Klypin}, {Zhao} \& {Somerville}}{{Klypin}
  et~al.}{2002}]{KZS02}
{Klypin} A.,  {Zhao} H.,    {Somerville} R.~S.,  2002, \apj, 573, 597

\bibitem[\protect\citeauthoryear{{Kormendy}, {Fischer}, {Cornell} \&
  {Bender}}{{Kormendy} et~al.}{2007}]{Kor07}
{Kormendy} J.,  {Fischer} D.~B.,  {Cornell} M.~E.,    {Bender} R.,  2007,
  submitted to \apj

\bibitem[\protect\citeauthoryear{{Korycansky}, {Laughlin} \&
  {Adams}}{{Korycansky} et~al.}{2001}]{Kory01}
{Korycansky} D.~G.,  {Laughlin} G.,    {Adams} F.~C.,  2001, \apss, 275, 349

\bibitem[\protect\citeauthoryear{{Li} \& {White}}{{Li} \& {White}}{2007}]{LW07}
{Li} Y.-S.,  {White} S.~D.~M.,  2007, MNRAS accepted (astro-ph/0710.3740), 710

\bibitem[\protect\citeauthoryear{{Loeb}}{{Loeb}}{2002}]{Loeb02}
{Loeb} A.,  2002, \prd, 65, 047301

\bibitem[\protect\citeauthoryear{{Loeb}, {Reid}, {Brunthaler} \&
  {Falcke}}{{Loeb} et~al.}{2005}]{Loeb05}
{Loeb} A.,  {Reid} M.~J.,  {Brunthaler} A.,    {Falcke} H.,  2005, \apj, 633,
  894

\bibitem[\protect\citeauthoryear{{McConnachie}, {Irwin}, {Ferguson}, {Ibata},
  {Lewis} \& {Tanvir}}{{McConnachie} et~al.}{2005}]{McC05}
{McConnachie} A.~W.,  {Irwin} M.~J.,  {Ferguson} A.~M.~N.,  {Ibata} R.~A.,
  {Lewis} G.~F.,    {Tanvir} N.,  2005, \mnras, 356, 979

\bibitem[\protect\citeauthoryear{{McConnachie}, {Venn}, {Irwin}, {Young} \&
  {Geehan}}{{McConnachie} et~al.}{2007}]{McC07}
{McConnachie} A.~W.,  {Venn} K.~A.,  {Irwin} M.~J.,  {Young} L.~M.,    {Geehan}
  J.~J.,  2007, \apjl, 671, L33

\bibitem[\protect\citeauthoryear{{Mihos} \& {Hernquist}}{{Mihos} \&
  {Hernquist}}{1994}]{MH94dsc}
{Mihos} J.~C.,  {Hernquist} L.,  1994, ApJL, 437, L47

\bibitem[\protect\citeauthoryear{{Mihos} \& {Hernquist}}{{Mihos} \&
  {Hernquist}}{1996}]{MH96}
{Mihos} J.~C.,  {Hernquist} L.,  1996, \apj, 464, 641

\bibitem[\protect\citeauthoryear{{Mo}, {Mao} \& {White}}{{Mo}
  et~al.}{1998}]{MMW98}
{Mo} H.~J.,  {Mao} S.,    {White} S.~D.~M.,  1998, \mnras, 295, 319

\bibitem[\protect\citeauthoryear{{Naab}, {Jesseit} \& {Burkert}}{{Naab}
  et~al.}{2006}]{NJB06}
{Naab} T.,  {Jesseit} R.,    {Burkert} A.,  2006, \mnras, 372, 839

\bibitem[\protect\citeauthoryear{{Naab} \& {Ostriker}}{{Naab} \&
  {Ostriker}}{2007}]{NO07}
{Naab} T.,  {Ostriker} J.~P.,  2007, ArXiv Astrophysics e-prints

\bibitem[\protect\citeauthoryear{{Nagamine} \& {Loeb}}{{Nagamine} \&
  {Loeb}}{2003}]{NL03}
{Nagamine} K.,  {Loeb} A.,  2003, New Astronomy, 8, 439

\bibitem[\protect\citeauthoryear{{Nagamine} \& {Loeb}}{{Nagamine} \&
  {Loeb}}{2004}]{NL04}
{Nagamine} K.,  {Loeb} A.,  2004, New Astronomy, 9, 573

\bibitem[\protect\citeauthoryear{{Navarro}, {Frenk} \& {White}}{{Navarro}
  et~al.}{1996}]{NFW96}
{Navarro} J.~F.,  {Frenk} C.~S.,    {White} S.~D.~M.,  1996, \apj, 462, 563

\bibitem[\protect\citeauthoryear{{Navarro}, {Frenk} \& {White}}{{Navarro}
  et~al.}{1997}]{NFW97}
{Navarro} J.~F.,  {Frenk} C.~S.,    {White} S.~D.~M.,  1997, \apj, 490, 493

\bibitem[\protect\citeauthoryear{{Nicastro}, {Zezas}, {Drake}, {Elvis},
  {Fiore}, {Fruscione}, {Marengo}, {Mathur} \& {Bianchi}}{{Nicastro}
  et~al.}{2002}]{Nic02}
{Nicastro} F.,  {Zezas} A.,  {Drake} J.,  {Elvis} M.,  {Fiore} F.,  {Fruscione}
  A.,  {Marengo} M.,  {Mathur} S.,    {Bianchi} S.,  2002, \apj, 573, 157

\bibitem[\protect\citeauthoryear{{Nicastro}, {Zezas}, {Elvis}, {Mathur},
  {Fiore}, {Cecchi-Pestellini}, {Burke}, {Drake} \& {Casella}}{{Nicastro}
  et~al.}{2003}]{Nic03}
{Nicastro} F.,  {Zezas} A.,  {Elvis} M.,  {Mathur} S.,  {Fiore} F.,
  {Cecchi-Pestellini} C.,  {Burke} D.,  {Drake} J.,    {Casella} P.,  2003,
  \nat, 421, 719

\bibitem[\protect\citeauthoryear{{Osone}, {Makishima}, {Matsuzaki}, {Ishisaki}
  \& {Fukazawa}}{{Osone} et~al.}{2002}]{Oso02}
{Osone} S.,  {Makishima} K.,  {Matsuzaki} K.,  {Ishisaki} Y.,    {Fukazawa} Y.,
   2002, \pasj, 54, 387

\bibitem[\protect\citeauthoryear{{Pahre}, {de Carvalho} \&
  {Djorgovski}}{{Pahre} et~al.}{1998}]{Pah98}
{Pahre} M.~A.,  {de Carvalho} R.~R.,    {Djorgovski} S.~G.,  1998, \aj, 116,
  1606

\bibitem[\protect\citeauthoryear{{Peebles}}{{Peebles}}{1994}]{P94}
{Peebles} P.~J.~E.,  1994, \apj, 429, 43

\bibitem[\protect\citeauthoryear{{Peebles}, {Melott}, {Holmes} \&
  {Jiang}}{{Peebles} et~al.}{1989}]{Peeb89}
{Peebles} P.~J.~E.,  {Melott} A.~L.,  {Holmes} M.~R.,    {Jiang} L.~R.,  1989,
  \apj, 345, 108

\bibitem[\protect\citeauthoryear{{Peebles}, {Phelps}, {Shaya} \&
  {Tully}}{{Peebles} et~al.}{2001}]{P01}
{Peebles} P.~J.~E.,  {Phelps} S.~D.,  {Shaya} E.~J.,    {Tully} R.~B.,  2001,
  \apj, 554, 104

\bibitem[\protect\citeauthoryear{{Raychaudhury} \&
  {Lynden-Bell}}{{Raychaudhury} \& {Lynden-Bell}}{1989}]{RLB89}
{Raychaudhury} S.,  {Lynden-Bell} D.,  1989, \mnras, 240, 195

\bibitem[\protect\citeauthoryear{{Ribas}, {Jordi}, {Vilardell}, {Fitzpatrick},
  {Hilditch} \& {Guinan}}{{Ribas} et~al.}{2005}]{Rib05}
{Ribas} I.,  {Jordi} C.,  {Vilardell} F.,  {Fitzpatrick} E.~L.,  {Hilditch}
  R.~W.,    {Guinan} E.~F.,  2005, \apjl, 635, L37

\bibitem[\protect\citeauthoryear{{Rothberg} \& {Joseph}}{{Rothberg} \&
  {Joseph}}{2004}]{RJ04}
{Rothberg} B.,  {Joseph} R.~D.,  2004, \aj, 128, 2098

\bibitem[\protect\citeauthoryear{{Sackmann}, {Boothroyd} \&
  {Kraemer}}{{Sackmann} et~al.}{1993}]{SBK93}
{Sackmann} I.-J.,  {Boothroyd} A.~I.,    {Kraemer} K.~E.,  1993, \apj, 418, 457

\bibitem[\protect\citeauthoryear{{Sanders} \& {Mirabel}}{{Sanders} \&
  {Mirabel}}{1996}]{SM96}
{Sanders} D.~B.,  {Mirabel} I.~F.,  1996, \araa, 34, 749

\bibitem[\protect\citeauthoryear{{Savage}, {Sembach}, {Wakker}, {Richter},
  {Meade}, {Jenkins}, {Shull}, {Moos} \& {Sonneborn}}{{Savage}
  et~al.}{2003}]{Sav03}
{Savage} B.~D.,  {Sembach} K.~R.,  {Wakker} B.~P.,  {Richter} P.,  {Meade} M.,
  {Jenkins} E.~B.,  {Shull} J.~M.,  {Moos} H.~W.,    {Sonneborn} G.,  2003,
  \apjs, 146, 125

\bibitem[\protect\citeauthoryear{{Sawa} \& {Fujimoto}}{{Sawa} \&
  {Fujimoto}}{2005}]{SF05}
{Sawa} T.,  {Fujimoto} M.,  2005, \pasj, 57, 429

\bibitem[\protect\citeauthoryear{{Seigar}, {Barth} \& {Bullock}}{{Seigar}
  et~al.}{2006}]{SBB07}
{Seigar} M.~S.,  {Barth} A.~J.,    {Bullock} J.~S.,  2006, ApJ submitted
  (astro-ph/0612228)

\bibitem[\protect\citeauthoryear{{Sembach}, {Wakker}, {Savage}, {Richter},
  {Meade}, {Shull}, {Jenkins}, {Sonneborn} \& {Moos}}{{Sembach}
  et~al.}{2003}]{Sem03}
{Sembach} K.~R.,  {Wakker} B.~P.,  {Savage} B.~D.,  {Richter} P.,  {Meade} M.,
  {Shull} J.~M.,  {Jenkins} E.~B.,  {Sonneborn} G.,    {Moos} H.~W.,  2003,
  \apjs, 146, 165

\bibitem[\protect\citeauthoryear{{Shen}, {Mo}, {White}, {Blanton}, {Kauffmann},
  {Voges}, {Brinkmann} \& {Csabai}}{{Shen} et~al.}{2003}]{Shen03}
{Shen} S.,  {Mo} H.~J.,  {White} S.~D.~M.,  {Blanton} M.~R.,  {Kauffmann} G.,
  {Voges} W.,  {Brinkmann} J.,    {Csabai} I.,  2003, \mnras, 343, 978

\bibitem[\protect\citeauthoryear{{Spergel}, {Verde}, {Peiris}, {Komatsu},
  {Nolta}, {Bennett}, {Halpern}, {Hinshaw}, {Jarosik}, {Kogut}, {Limon},
  {Meyer}, {Page}, {Tucker}, {Weiland}, {Wollack} \& {Wright}}{{Spergel}
  et~al.}{2003}]{Sp03}
{Spergel} D.~N.,  {Verde} L.,  {Peiris} H.~V.,  {Komatsu} E.,  {Nolta} M.~R.,
  {Bennett} C.~L.,  {Halpern} M.,  {Hinshaw} G.,  {Jarosik} N.,  {Kogut} A.,
  {Limon} M.,  {Meyer} S.~S.,  {Page} L.,  {Tucker} G.~S.,  {Weiland} J.~L.,
  {Wollack} E.,    {Wright} E.~L.,  2003, \apjs, 148, 175

\bibitem[\protect\citeauthoryear{{Springel}}{{Springel}}{2000}]{Sp00}
{Springel} V.,  2000, \mnras, 312, 859

\bibitem[\protect\citeauthoryear{{Springel}}{{Springel}}{2005}]{SpGad2}
{Springel} V.,  2005, \mnras, 364, 1105

\bibitem[\protect\citeauthoryear{{Springel}, {Di Matteo} \&
  {Hernquist}}{{Springel} et~al.}{2005}]{SdMH05}
{Springel} V.,  {Di Matteo} T.,    {Hernquist} L.,  2005, \mnras, 361, 776

\bibitem[\protect\citeauthoryear{{Springel} \& {Hernquist}}{{Springel} \&
  {Hernquist}}{2002}]{SHEnt}
{Springel} V.,  {Hernquist} L.,  2002, \mnras, 333, 649

\bibitem[\protect\citeauthoryear{{Springel} \& {Hernquist}}{{Springel} \&
  {Hernquist}}{2003}]{SH03}
{Springel} V.,  {Hernquist} L.,  2003, \mnras, 339, 289

\bibitem[\protect\citeauthoryear{{Springel} \& {White}}{{Springel} \&
  {White}}{1999}]{SW99}
{Springel} V.,  {White} S.~D.~M.,  1999, \mnras, 307, 162

\bibitem[\protect\citeauthoryear{{Stinson}, {Seth}, {Katz}, {Wadsley},
  {Governato} \& {Quinn}}{{Stinson} et~al.}{2006}]{Stin06}
{Stinson} G.,  {Seth} A.,  {Katz} N.,  {Wadsley} J.,  {Governato} F.,
  {Quinn} T.,  2006, \mnras, 373, 1074

\bibitem[\protect\citeauthoryear{{Tegmark}}{{Tegmark}}{2006}]{Teg06short}
{Tegmark} M. e.~a.,  2006, \prd, 74, 123507

\bibitem[\protect\citeauthoryear{{Toomre}}{{Toomre}}{1977}]{T77}
{Toomre} A.,  1977, in Evolution of Galaxies and Stellar Populations {Mergers
  and Some Consequences}.
p. p.401

\bibitem[\protect\citeauthoryear{{Toomre} \& {Toomre}}{{Toomre} \&
  {Toomre}}{1972}]{TT72}
{Toomre} A.,  {Toomre} J.,  1972, \apj, 178, 623

\bibitem[\protect\citeauthoryear{{Udry}, {Bonfils}, {Delfosse}, {Forveille},
  {Mayor}, {Perrier}, {Bouchy}, {Lovis}, {Pepe}, {Queloz} \& {Bertaux}}{{Udry}
  et~al.}{2007}]{Udry07}
{Udry} S.,  {Bonfils} X.,  {Delfosse} X.,  {Forveille} T.,  {Mayor} M.,
  {Perrier} C.,  {Bouchy} F.,  {Lovis} C.,  {Pepe} F.,  {Queloz} D.,
  {Bertaux} J.~.,  2007, A\&A submmitted (astro-ph/0704.3841), 704

\bibitem[\protect\citeauthoryear{{Valtonen}, {Byrd}, {McCall} \&
  {Innanen}}{{Valtonen} et~al.}{1993}]{Val93}
{Valtonen} M.~J.,  {Byrd} G.~G.,  {McCall} M.~L.,    {Innanen} K.~A.,  1993,
  \aj, 105, 886

\bibitem[\protect\citeauthoryear{{van der Marel} \& {Guhathakurta}}{{van der
  Marel} \& {Guhathakurta}}{2007}]{vdM07}
{van der Marel} R.~P.,  {Guhathakurta} P.,  2007, ApJ submitted
  (astro-ph/0709.3747), 709

\bibitem[\protect\citeauthoryear{{Widrow} \& {Dubinski}}{{Widrow} \&
  {Dubinski}}{2005}]{WD05}
{Widrow} L.~M.,  {Dubinski} J.,  2005, \apj, 631, 838

\bibitem[\protect\citeauthoryear{{Willman}, {Dalcanton}, {Martinez-Delgado},
  {West}, {Blanton}, {Hogg}, {Barentine}, {Brewington}, {Harvanek}, {Kleinman},
  {Krzesinski}, {Long}, {Neilsen} Jr., {Nitta} \& {Snedden}}{{Willman}
  et~al.}{2005}]{Will05}
{Willman} B.,  {Dalcanton} J.~J.,  {Martinez-Delgado} D.,  {West} A.~A.,
  {Blanton} M.~R.,  {Hogg} D.~W.,  {Barentine} J.~C.,  {Brewington} H.~J.,
  {Harvanek} M.,  {Kleinman} S.~J.,  {Krzesinski} J.,  {Long} D.,  {Neilsen}
  Jr. E.~H.,  {Nitta} A.,    {Snedden} S.~A.,  2005, \apjl, 626, L85

\bibitem[\protect\citeauthoryear{{Wyse}}{{Wyse}}{2007}]{Wyse07}
{Wyse} R.~F.~G.,  2007, in IAU Symposium Vol.~77 of IAU Symposium, {Lessons
  from Surveys of The Galaxy}.
pp 1036--1046

\end{thebibliography}

\end{document}